\documentclass{aa}  

\usepackage{graphicx}
\usepackage{caption}
\usepackage{subcaption}
\usepackage{comment}
\usepackage{txfonts}
\usepackage[switch]{lineno} 
\usepackage[bookmarks=false,colorlinks=true,linkcolor=black,citecolor=cyan,filecolor=black,urlcolor=cyan]{hyperref}
\usepackage{xcolor}

\begin{document} 


   \title{The hot circumgalactic medium in the eROSITA All-Sky Survey}
   \subtitle{I. X-ray surface brightness profiles}
   \author{Yi Zhang \inst{1}\fnmsep\thanks{yizhang@mpe.mpg.de} \and
    Johan Comparat\inst{1} \and Gabriele Ponti\inst{2,1} \and Andrea Merloni\inst{1}
    \and Kirpal Nandra\inst{1}
    \and Frank Haberl\inst{1}
    \and Nicola Locatelli\inst{2}
    \and Xiaoyuan Zhang\inst{1}
    \and Jeremy Sanders\inst{1}
    \and Xueying Zheng\inst{1}
    \and Ang Liu\inst{1}
    \and Paola Popesso\inst{3}
    \and Teng Liu\inst{4,5}
    \and Nhut Truong \inst{6,7}
    \and Annalisa Pillepich \inst{8}
    \and Peter Predehl\inst{1}
    \and Mara Salvato\inst{1}
    \and Soumya Shreeram\inst{1}
    \and Michael C. H. Yeung\inst{1}
    \and Qingling Ni\inst{1}
          }

   \institute{Max-Planck-Institut für extraterrestrische Physik (MPE), Gießenbachstraße 1, D-85748 Garching bei München, Germany
   \and
    INAF-Osservatorio Astronomico di Brera, Via E. Bianchi 46, I-23807 Merate (LC), Italy 
    \and 
    European Southern Observatory, Karl Schwarzschildstrasse 2, D-85748 Garching bei München, Germany 
    \and
    Department of Astronomy, University of Science and Technology of China, Hefei 230026, China 
    \and 
    School of Astronomy and Space Science, University of Science and Technology of China, Hefei 230026, China
    \and
    NASA Goddard Space Flight Center, Greenbelt, MD 20771, USA 
    \and 
     Center for Space Sciences and Technology, University of Maryland, 1000 Hilltop Circle, Baltimore, MD 21250, USA 
    \and
    Max-Planck-Institut für Astronomie, Königstuhl 17, 69117 Heidelberg, Germany 
             }
   \date{ \today }

 
 \abstract
   {The circumgalactic medium (CGM) provides the material needed for galaxy formation and influences galaxy evolution.
   The hot ($T>10^6K$) CGM is poorly detected around galaxies with stellar masses ($M_*$) lower than $3\times10^{11}M_\odot$ due to the low surface brightness.
   }
   {We aim to detect the X-ray emission from the hot CGM around Milky Way-mass (MW-mass, $\log (M_*/M_\odot)=10.5-11.0$) and M31-mass ($\log (M_*/M_\odot)=11.0-11.25$) galaxies, in addition to measuring the X-ray surface brightness profile of the hot CGM.}
   {We applied a stacking technique to gain enough statistics to detect the hot CGM. We used the X-ray data from the first four SRG/eROSITA All-Sky Surveys (eRASS:4). We discussed how the satellite galaxies could bias the stacking and the method we used to carefully build the central galaxy samples. Based on the SDSS spectroscopic survey and halo-based group finder algorithm, we selected central galaxies with spectroscopic redshifts of $z_{\rm spec}<0.2$ and stellar masses of $10.0<\log(M_*/M_\odot)<11.5$ (85,222 galaxies) -- or halo masses of $11.5<\log(M_{\rm 200m}/M_\odot)<14.0$ (125,512 galaxies). By stacking the X-ray emission around galaxies, we obtained the mean X-ray surface brightness profiles. We masked the detected X-ray point sources and carefully modeled the X-ray emission from the unresolved active galactic nuclei (AGN) and X-ray binaries (XRB) to obtain the X-ray emission from the hot CGM.}
   {We measured the X-ray surface brightness profiles for central galaxies of $\log(M_*/M_\odot)>10.0$ or $\log(M_{\rm 200m}/M_\odot)>11.5$. We detected the X-ray emission around MW-mass and more massive central galaxies extending up to the virial radius ($R_{\rm vir}$). The signal-to-noise ratio (S/N) of the extended emission around MW-mass (M31-mass) galaxy is about $3.1\sigma$ ($4.7\sigma$) within $R_{\rm vir}$. We used a $\beta$ model to describe the X-ray surface brightness profile of the hot CGM ($S_{\rm X,CGM}$). We obtained a central surface brightness of $\log(S_{\rm X,0}[\rm erg\,s^{-1}\,kpc^{-2}])= 36.7^{+1.4}_{-0.4}\,(37.1^{+1.5}_{-0.4})$ and $\beta =0.43^{+0.10}_{-0.06}\,(0.37^{+0.04}_{-0.02})$ for MW-mass (M31-mass) galaxies. For galaxies with $\log(M_{\rm 200m}/M_\odot)>12.5$, the extended X-ray emission is detected with S/N $>2.8 \sigma$ and the $S_{\rm X,CGM}$ can be described by a $\beta$ model with $\beta\approx0.4$ and $\log(S_{\rm X,0}[\rm erg\,s^{-1}\,kpc^{-2}]) > 37.2$.
   We estimated the baryon budget of the hot CGM and obtained a value that is lower than the prediction of $\Lambda$CDM cosmology, indicating significant gas depletion in these halos. We extrapolated the hot CGM profile measured within $R_{\rm vir}$ to larger radii and found that within $\approx 3 R_{\rm vir}$, the baryon budget is close to the $\Lambda$CDM cosmology prediction.}
   {We measured the extended X-ray emission from representative populations of central galaxies around and above MW-mass out to $R_{\rm vir}$. Our results set a firm footing for the presence of the hot CGM around such galaxies. These measurements constitute a new benchmark for galaxy evolution models and possible implementations of feedback processes therein.}


   \maketitle

%
\section{Introduction}

The circumgalactic medium (CGM) is the gas reservoir of a galaxy and it is closely related to galaxy evolution processes. Understanding the properties of the CGM is essential to the study of galaxy formation and evolution, for example, the strength of feedback processes and how they modulate the star formation and gas activity \citep{TumlinsonPeeplesWerk_2017ARAA..55..389T, NaabOstriker_2017ARAA..55...59N, Truong2020, Truong2021a, EckertGaspariGastaldello_2021Univ....7..142E,OppenheimerBabulBahe_2021Univ....7..209O}.
The observed CGM has multiple phases. Its temperature ($T$) ranges from $<10^4$ K to $>10^6$ K. 
The cold ($T<10^4$ K), cool ($T=10^4-10^5$ K) and warm ($T=10^5-10^6$ K) phases of CGM have been well studied \citep{Prochaska2011,Tumlinson2011,Putman2012,Werk2014,Werk2016,Prochaska2017}. The next vital target is to attain a complete picture of the relationship between the galaxy and its hot ($T>10^6 K$) phase of the CGM. 

The hot phase of the CGM is the dominant mass component of the CGM baryon budget \citep{TumlinsonPeeplesWerk_2017ARAA..55..389T}. This hot CGM can be heated and shaped by the gravitational accretion shock and feedback processes, for example, active galactic nucleus (AGN) and stellar feedback.
It emits X-rays via collisionally ionized gas emission lines, recombination lines and bremsstrahlung. 
Pointed observations using {\textit{Chandra}} or {\textit{XMM-Newton}} revealed the hot gas properties of a few nearby galaxies within a distance of 50 Mpc \citep{Strickland2004, Tullmann2006, Wang2010, Dai2012, Li2013b, Bogdan2013, Bogdan2015, Anderson2016, Li2016}. 
The hot gas surveys of {\textit{Chandra}} and {\textit{XMM-Newton}} focus on the inner hot CGM (or the "corona") with scale heights of about 1-30 kpc (median of 5 kpc), where feedback processes dominate over gravitation. 
With their spatial resolution, {\textit{XMM-Newton}} and especially {\textit{Chandra}} resolve the stellar structures distributed in the disk of nearby galaxies. Still, the availability of galaxies with the detectable extended emission of the CGM is limited in number and extension.

For the outer hot CGM, where virialized gas dominates, to heat the gas to X-ray emitting temperature over the radiation cooling curve through gravitation, $M_{\rm halo}$ needs to be high enough (typically >$10^{12.7} M_\odot$). This corresponds to massive galaxies with about $M_*>10^{11}M_\odot$  \citep{Keres2009a,Keres2009b,Voort2016,Li2016,Liu2022}. 
For these reasons, although the hot gas in galaxy groups and clusters has been detected and studied well \citep{Pratt2019, WalkerSimionescuNagai_2019SSRv..215....7W}, very few detections of the outer hot CGM (out to 70 kpc) from galaxies have been reported \citep{Dai2012, Bogdan2013, Bogdan2015, Anderson2016, Li2016, Das2020}. 

To chart the CGM outer emission beyond the Local Universe, we use stacking techniques to detect its faint emission, provided large enough volumes are surveyed. 
By stacking about 250,000 "locally brightest galaxies" selected from the Sloan Digital Sky Survey (SDSS) and using data from ROSAT All-Sky Survey, \citet{Anderson2015} detected hot gas emission around $M_*>10^{10.8}M_\odot$ galaxies. 
In addition, X-ray observations and studies of the CGM with the thermal Sunyaev–Zeldovich (tSZ) effect have progressed in recent years; for example, with the Atacama Cosmology Telescope \citep[ACT,][]{AiolaCalabreseMaurin_2020JCAP...12..047A}. The thermal energy density profile of the CGM is measured around $M_*>10^{10.6}M_\odot$ galaxies by cross-correlating and stacking the WISE and SuperCOSmos photometric redshift galaxy sample with ACT and Planck \citep{BilickiPeacockJarrett_2016ApJS..225....5B, DasChiangMathur_2023ApJ...951..125D}.

X-ray stacking experiments are now dramatically progressing thanks to the extended ROentgen survey with an Imaging Telescope Array (eROSITA) on board the Spektrum-Roentgen-Gamma (SRG) orbital observatory. 
Launched in 2019, eROSITA is a sensitive X-ray telescope performing all-sky surveys with a wide-field focusing telescope \citep{Merloni2012, Predehl2021, SunyaevArefievBabyshkin_2021A&A...656A.132S,Merloni2024}. 
The sensitivity of eROSITA is maximal in the soft energy band, namely below 2.3 keV, which makes it suitable for studying the million-degrees hot CGM emission. 
Recently, with eROSITA and its PV/eFEDS observations covering 140 $\rm deg^2$, \citet{Comparat2022} stacked $\sim$16,000 central galaxies selected from the Galaxy And Mass Assembly (GAMA) spectroscopic galaxy survey; 
later, \citet{Chada2022} stacked $\sim$1,600 massive galaxies selected from SDSS in the same eFEDS field. 
Both works detected and measured weak X-ray emission from the outer CGM of those galaxies by stacking star-forming and quiescent galaxies separately. The difference in the sample selection has led to different interpretations of the results. 

At the time of writing, four of the planned eight all-sky surveys (eROSITA all-sky survey, eRASS:4), each lasting six months, have been completed. The X-ray data currently available for CGM stacking are about 40 times larger than what was available for the eFEDS region (100 times more area covered with about 40\% the exposure time). 
In this work and the accompanying papers, we stack the eRASS:4 X-ray data and a set of galaxy samples with high statistical completeness.
We study the X-ray surface brightness\footnote{In this work and companion papers, we define the surface brightness as the luminosity of the source per square kiloparsec, see Eq.~\ref{Equ_Sx}.} profile of the hot CGM. 
In the companion papers based on the same samples, we investigate (i) the scaling relations between X-ray luminosity, stellar mass, and halo mass \citep{Zhang2024relation}; (ii) the trends as a function of specific star formation rate (Zhang et al., in prep); and (iii) the possible azimuthal dependence of the hot CGM emission (Zhang et al., in prep).

We organize this paper as follows. 
We introduce the eROSITA X-ray data reduction and stacking method in Sect.~\ref{Sec_data}. 
We define the galaxy samples in Sect.~\ref{Sec_galaxy}. 
We describe the models (AGN, XRB) and mock galaxy catalogs in Sect.~\ref{Sec_model}. 
We present the mean X-ray surface brightness profiles of the complete galaxy population in Sect.~\ref{Sec_profilefullsat}. 
We show the X-ray surface brightness profiles of central galaxies and their hot CGM in Sect.~\ref{Sec_profilegal}. We estimate the baryon budget for the central galaxies in Sect.~\ref{Sec_baryon}. 
Throughout the paper, we use \citet{Planck2020} cosmological parameters: $H_0 = 67.74\ \rm km\,s^{-1}\,Mpc^{-1}$ and $\Omega_{\rm m} = 0.3089$. All instances of $\log$ in this work are in base 10.

\section{X-ray data reduction and stacking method}
\label{Sec_data}

In this section we describe the reduction of the X-ray data (Sect.~\ref{Sec_Xray}), stacking method (Sect.~\ref{Sec_method}), and calculation of the uncertainties (Sect.~\ref{Sec_error}), along with the PSF and $\beta$ models (Sect.~\ref{Sec_PSF}).

\subsection{X-ray data reduction}\label{Sec_Xray}
We used the eRASS:4 data in the western Galactic hemisphere ($179.9442^\circ<l\leqslant 359.9442^\circ$).
For the analysis, we used the eROSITA Science Analysis Software System \citep[eSASS;][]{Brunner2022} data products (version 020) \citep{Merloni2024}. 
We staged the data in 2439 overlapping sky tiles, each covering $3.6^\circ \times 3.6^\circ$. 
For each sky tile, we had energy-calibrated event files, vignetting-corrected mean exposure map in three bands (0.3--0.6 keV, 0.6--1.0 keV, 1.0--2.0 keV), and a catalog of detected sources in the 0.2--2.3\,keV band down to a low detection likelihood of \texttt{DET\_LIKE\_0}$ > 5$. 
We used the default flag and keep all events with valid patterns. 
eROSITA has seven telescope modules (TMs); two TMs suffer from light leaks \citep[TM5 and 7;][]{Predehl2021} and have a higher background. We used all seven telescope modules (TMs) in our analysis. We have verified that the inclusion of the two light-leak cameras (TM5 and 7) does not affect our results\footnote{We compared the stacking results when including or excluding the TM5 and TM7; without these two TMs, the background is lower while the statistics are worse. Our stacking results are independent of their inclusion.}. 
The vignetted exposure time of eRASS:4, in the 0.5--2~keV band, ranges from 300s to 10,000s, with a median value of 550\,s.

To account for the absorption from the Galactic neutral hydrogen, we masked the sky area where $N_{\rm H}>10^{21}\,\rm cm^{-2}$ as traced by the integrated HI4PI column density map \citep{HI4PI2016}. 
The dust extinction and reddening in the Milky Way could cause inaccuracies in galaxy parameters. Using the thermal dust emission model from Planck observation \citep{Planck2013} as a proxy, we kept only the sky regions with $E(B-V)<0.1$. 

The source detection pipeline of eROSITA may fail around bright sources (for example, Sco X-1 and the Virgo cluster) \citep{Merloni2024}.
We masked these overdense source-detection regions where potentially spurious sources are clustered (see the masking procedure in \citet{Merloni2024}).
After applying the mask, about 11,000 deg$^2$ covered by X-ray observations ($A_{\rm X}$) remained.

\subsection{Stacking method} \label{Sec_method}

We adopted the stacking method of \citet{Comparat2022}. For each galaxy in the sample, we created an event cube by retrieving events within 3 Mpc around the galaxy ($\rm CUBE_{event}$). Each event cube contains the following information for each event in the cube: the position (RA, DEC), the angular separation and physical distance\footnote{The physical (proper) distance between events and galaxies are derived according to the redshift of the galaxy and angular separation.} ($R_{\rm rad}$ and $R_{\rm kpc}$) to the associated galaxy, the exposure time, $t_{\rm exp}$, the observed energy, $E_{\rm obs}$, the corresponding rest frame energy, $E_{\rm rest}=E_{\rm obs}\times(1+z),$ and the effective area, $A_{\rm eff}$, of the telescope at the observed energy. In the case of masking sources detected by eRASS:4, we took the masking radius of each source conservatively as $1.4\times R_{\rm src}$ to avoid residual emission, where $R_{\rm src}$ is the source radius derived from \texttt{srctool} \citep[see Appendix A of][]{Comparat2023}. A mask flag is assigned to an event if it falls in these masked regions.

Stacking a galaxy sample consists of merging the event cubes for all galaxies in the sample and extracting a summary statistic. 
Each event is assigned a weight: 
\begin{equation}
    W_e=E_{\rm obs} \frac{1}{t_{\rm exp}}\frac{f_{\rm nh} A_{\rm corr}\,4\pi D_L^2 }{ A_{\rm eff}}\ [\rm erg\,s^{-1}\,Mpc^2\,cm^{-2}], 
\end{equation}
where $D_{\rm L}$ is the luminosity distance to the galaxy. Also,
$f_{\rm nh}$ is a function that corrects the effects of absorption on the soft X-ray photons as a function of the $E_{\rm obs}$ and $N_{\rm H}$ (taken from \citet{HI4PI2016}). We obtain $f_{\rm nh}$ using the TBABS absorption model \citep{Wilms2000}. $A_{\rm corr}$ is an area correction factor that accounts for the area loss due to masking X-ray sources. 

The X-ray surface brightness profile around galaxies ($S_{\rm X, gal}$) is calculated by selecting the events located in radial bins ($\rm r_0<R_{kpc}<r_1$) and summing them up as:\ 
\begin{equation}\label{Equ_Sx}
    S_{\rm X,gal}= \sum_{r_0}^{r_1} \frac{ W_{\rm e} }{A_{\rm shell} N_{\rm g}}\ [\rm erg\,s^{-1}\,kpc^{-2}],
\end{equation}
where $A_{\rm shell}$ is the area of each radial bin and $N_{\rm g}$ is the number of galaxies stacked\footnote{The radial bin boundaries used are (0, 10, 30, 50, 75, 100, 150, 200, 250, 300, 350, 400, 500, 600, 700, 800, 900, 1000, 1100, 1200, 1300, 1400, 1600, 1800, 2000, 2200, 2400, 2600, 2800, 3000) kpc.}.

The background surface brightness ($S_{\rm X,bg}$) is taken as the minimum value of $S_{\rm X,gal}$ ($S_{\rm X,min}$) beyond $R_{\rm 500c}$\footnote{$R_{\rm 500c}$ is the radius where the density is 500 times the background universe critical density.}. 
We detail the background estimation procedure validation in Appendix.~\ref{Apd_bg}. 
Throughout this work, we use the energy bin $0.5-2$ keV (in rest frame) to calculate the X-ray surface brightness.

\subsection{Uncertainty calculation}\label{Sec_error}

The uncertainty on $S_{\rm X, gal}$ contains two components: the Poisson error from the data and the uncertainties from the stacking process.
We estimated the Poisson error of the X-ray surface brightness profile by $1/\sqrt{n_{\rm total}}$, where $n_{\rm total}$ is the total event numbers in the corresponding radial bin. 

The uncertainties from the stacking process were computed using the Jackknife re-sampling method \citep{Andraejackknife2010, McIntoshjackknife2016} to assess the X-ray property variation of the stacked galaxy population. For a galaxy sample containing $N_{\rm gal}$ galaxies, we randomly selected $0.9N_{\rm gal}$ galaxies from the sample, then stacked the $0.9N_{\rm gal}$ galaxies and obtain the mean $S_{\rm X, gal}$ of them. We repeated the above procedure 50 times and retrieved the standard deviation of the 50 mean $S_{\rm X, gal}$ as the uncertainty from the stacking process.

For reference, the uncertainties in our study are dominated by the standard deviation of the mean $S_{\rm X, gal}$, which reflects the intrinsic scatter in the X-ray surface brightness of the stacked galaxy population. 
It is therefore possible to obtain smaller overall uncertainties of $S_{\rm X, gal}$ using a masked sample despite its lower photon statistics, since masking significantly decreases the stacking uncertainties coming from point sources.

\subsection{Point spread function and $\beta$ models}\label{Sec_PSF}

We used the eROSITA point spread function (PSF) model to assess the extended nature of the observed surface brightness profile. We also used it to model the surface brightness profiles of point sources.
We used the eROSITA PSF model from Sanders et al. (in prep; also, \citet{Merloni2024}). 
The PSF is expressed in angular units. We use the redshifts of the galaxies to convert it to a physical scale (kpc). 
For each galaxy sample, we convolve the PSF profile (in kpc) with the redshift distribution\footnote{This assumes that the luminosity of the point sources in the stacked galaxies does not strongly depend on the redshift. The empirical XRB luminosity scales as $(1+z)^{3.79\pm0.12}$ \citep{Aird2017}, and the maximum redshift of our galaxy sample is 0.2. This suggests an evolution of less than two times. Detected X-ray point sources are masked, leaving only undetected ones in the stack. This justifies this assumption.}.
The extent of the X-ray emission was assessed by scaling the mean PSF profile to the central surface brightness observed to obtain a maximal PSF-like surface brightness profile. If excess emission (above the maximal PSF-like profile) is measured at large radii, then the X-ray emission must come from a source with an extension beyond that of the PSF, for example: the hot CGM.

 The $\beta$ model \citep{Cavaliere1976} is used to describe the X-ray profiles of the hot CGM:

\begin{equation}
\label{eq:Sbeta}
    S_{\rm X, \beta}=S_{\rm X,0}\left[1+\left(\frac{r}{r_{\rm c}}\right)^2\right]^{-3\beta+\frac{1}{2}}\ [\rm erg\,s^{-1}\,kpc^{-2}],
\end{equation} 

\noindent where $S_{\rm X,0}$ is the X-ray surface brightness at the galaxy center, $r$ is the distance from the center, and $r_{c}$ is the core radius. We convolved the $\beta$ profile with the PSF and built the likelihood function. We fit the PSF-convolved $\beta$ model to the observation using the Markov chain Monte Carlo (MCMC) method. We took the best-fit value as the median value of the possible parameters from the MCMC and the values at 16$^{\rm th}$ and 84$^{\rm th}$ percentiles from MCMC as the $1\sigma$ uncertainties \citep{emcee2010}.

\section{Galaxy samples selection} \label{Sec_galaxy}

We built our galaxy samples based on two criteria: (i)  completeness (>90\%) to reduce selection effects and (ii) size, whereby the larger, the better.

In this section, we calculate the maximum redshift limited by the angular resolution of eROSITA and the required galaxy survey depth to build a complete galaxy sample (Sect.~\ref{Sec_maxz}). 
Based on the two criteria above, we select the Sloan Digital Sky Survey data release 7 (SDSS DR7) spectroscopic galaxy catalog and DESI Legacy Survey data release 9 (DESI LS DR9) photometric galaxy catalog to build the galaxy samples (Sect.~\ref{Sec_choicecatalog}). 

We outline the galaxy samples and their purpose in Sect.~\ref{Sec_sample_sum}. We explain the constructions of samples in the following sections: We describe the FULL$_{\rm spec}$ and FULL$_{\rm phot}$ samples in Sect.~\ref{Sec_sdssls}. 
We discuss the connection between galaxies and their dark matter halos in Sect.~\ref{Sec_galhalo} and how the presence of satellite galaxies in the sample could bias the stacking result (Sect.~\ref{Sec_proj}). To minimize such bias, we define the central galaxy samples (CEN and CEN$_{\rm halo}$) based on the SDSS spectroscopic galaxy survey (Sect.~\ref{Sec_cen}) and an isolated galaxy sample based on LS DR9 photometric galaxy catalog (Appendix.~\ref{Sec_iso}).

\subsection{Considerations on the maximum redshift and galaxy survey depth}\label{Sec_maxz}

\begin{table*}[]
    \centering
    \caption{Constraints on the maximum redshifts of galaxies at different stellar masses to achieve $>95\%$ completeness.}
    \label{tab_zmax}
    \begin{tabular}{ccccccccccc}
\hline \hline
log$_{10}$(M$_*$) & log$_{10}$(M$_{\rm h}$) & R$_{\rm h}$ & $z_{\rm 2\times PSF}$ &$z_{\rm 3\times PSF}$& $r_{AB,\rm 2\times PSF}$ &$r_{AB,\rm 3\times PSF}$&$z_{\rm r_{AB}=17.77}$\\\
[$M_\odot$]&[$M_\odot$]&[kpc] \\
\hline
9    & 11.0 &  82  & 0.07&0.04 & 19.5 & 18.5&<0.02\\
9.5  & 11.3 & 102  & 0.09& 0.06& 19.4& 18.8&0.04\\
10   & 11.6 & 131  & 0.12& 0.075& 19.2&18.1&0.06\\
10.5 & 12.0 & 178  & 0.17&0.10 & 19.1&17.8&0.1\\
10.75 & 12.2 & 210  & 0.21& 0.13& 19.1&17.9&0.12\\
11.0 & 12.7 & 295& 0.33& 0.19& 19.4&18.3&0.15\\
11.25 & 13.3 & 480 &0.91&  0.37& 19.3&19.4&0.19\\
11.5 & 14.3 & 950 & >1& >1 & 19.2& 19.2&0.25\\
\hline
    \end{tabular}
\tablefoot{From left to right: Galaxy stellar mass, the corresponding average dark matter halo mass, the average halo radius, the redshift at which the projected halo radius equals two ($1'$) or three ($1.5'$) eROSITA PSF, the r-band survey depth required to achieve 95\% completeness within $z_{\rm 2\times PSF}$ or $z_{\rm 3\times PSF}$ (we take $z=0.4$ to calculate $r_{AB}$ if $z_{\rm 2\times PSF}$ or $z_{\rm 3\times PSF}$ exceeds 0.4), and redshift within which galaxy survey with r-band survey depth $r_{AB}=17.77$ achieves 95\% completeness.}
\end{table*}

To resolve the extended hot CGM, the PSF size of eROSITA limits the maximum redshift within which we should stack galaxies. 
With the stellar-to-halo mass relation from \citet{MosterNaabWhite_2013MNRAS.428.3121M}, we infer the average dark matter halo mass and virial radius for a galaxy of a given stellar mass. 
We convert the virial radius to angular scale using the angular diameter distance as a function of redshift. 
As a reference, we take the size of the eROSITA PSF as $\theta=30\arcsec$ (\citealt{Merloni2024}; Sanders et al. in prep.) and calculate at which redshift the virial radius of the halo hosting the galaxy is equal to two and three times this value (i.e. $1\arcmin$ and $1\farcm5$), as listed in Table~\ref{tab_zmax}. For example, galaxies with $\log (M_*)=10.75$ (similar to the Milky Way) have a resolved virial radius with two (three) eROSITA PSF up to $z=0.21$ ($z=0.13$). 
We then estimate the galaxy survey depth to ensure the galaxy sample's completeness is higher than 95\% (we take r-band magnitude $r_{\rm AB}$ to evaluate). For example, to resolve the virial halo of $\log (M_*)=10.75$ galaxies with two (three) eROSITA PSF, the galaxy survey needs to be deeper than $r_{\rm AB}=19.1 (17.9)$.

By stacking samples up to $z_{\rm 2\times PSF}$, we obtain higher statistics to measure the luminosity of the CGM, and the survey depth should be deeper than $r_{\rm AB}=19.1$. Stacking samples up to $z_{\rm 3\times PSF}$, we obtain a better spatial resolution to measure the surface brightness profile of the CGM, and the survey depth can be shallower ($r_{\rm AB}=17.9$).

\subsection{Choice of the parent galaxy surveys}\label{Sec_choicecatalog}

\begin{table*}[]
    \centering
    \caption{Galaxy catalogs used in this work.}
    \label{tab:catalogs}
    \begin{tabular}{ccccccc}
\hline \hline
Survey & $A_{\rm west}$ (deg$^2$) & Magnitude cut &$N_{\rm gal}$ & redshift & Reference  \\
\hline
SDSS (DR7)  &4010 &r<17.77&382428 & spectroscopic & \citet{Strauss2002} \\
DESI LS (DR9) & 9340 & r<23.4 & 2387058 & photometric  & \citet{Dey2019}\\

\hline
    \end{tabular}
\tablefoot{$A_{\rm west}$ is the overlap area of the galaxy survey and the western Galactic hemisphere. The survey depth of each catalog is shown in column three. $N_{\rm gal}$ is the total number of galaxies stacked within the maximum redshift in the western Galactic hemisphere at different mass bins defined in Sect.~\ref{Sec_maxz}. The calculation method of redshift and the relevant references are in the last two columns. }
\end{table*}

There are numerous photometric and spectroscopic surveys since decades, such as Two Micron All Sky Survey \citep[2MASS,][]{Skrutskie2006, Bilicki2021}, VISTA Hemisphere Survey \citep[VHS,][]{McMahon2013}, Wide-field Infrared Survey Explorer \citep[WISE,][]{WrightEisenhardtMainzer_2010AJ....140.1868W}, SDSS \citep[][]{AlmeidaAndersonArgudo-Fernandez_2023}, skymapper \citep[][]{Onken2019}, GAMA \citep[][]{Driver2022}, Dark Energy Survey \citep[DES,][]{Abbott2018}, Kilo Degree Survey \citep[KiDS,][]{deJong2017}, and DESI LS \citep[][]{Dey2019}. 

To make full use of the X-ray data at our disposal in the western Galactic hemisphere, the survey area is more important to us than depth. Therefore, we prefer wide-area surveys and do not consider deep but small fields such as HUDF \citep[Hubble Ultra Deep Field,][]{Beckwith2006}, CDFS \citep[\textit{Chandra} Deep Field South;][]{HSu2014}, and COSMOS \citep[Cosmic Evolution Survey,][]{Laigle2016}. Moreover, we prefer spectroscopic surveys that provide more accurate redshift measurements, stellar population, and galaxy classification. However, we also need to include photometric surveys to strive for large area and completeness before spectroscopic redshifts of bright galaxies ($r_{AB}<20$) from 4MOST and DESI become available over the full sky \citep{DESICollaborationAghamousaAguilar_2016arXiv161100036D,deJong2019, FinoguenovMerloniComparat_2019Msngr.175...39F}
\footnote{In the longer term, Rubin \citep{Ivezic_LSST_2019ApJ...873..111I} and Euclid \citep{LaureijsAmiauxArduini_2011arXiv1110.3193L} observatories will be excellent resources with exquisite depth and complete coverage of the southern hemisphere.}.

In this work, we built our galaxy samples based on two surveys: SDSS DR7 \citep{Strauss2002} and DESI LS DR9 \citep{Dey2019}, as listed in Table~\ref{tab:catalogs}. SDSS DR7 provides enough statistics to study the CGM around galaxies within $z=0.2$.
LS DR9 is currently the most adequate photometric survey and we took it to build the galaxy sample out to $z_{\rm 2\times PSF}$ and down to $\log(M_*)=9.5$, which provides the best statistics to detect the extended X-ray emission around galaxies.

\subsubsection{SDSS DR7}
The SDSS DR7 Main Galaxy Sample (MGS, $r_{\rm AB}<17.77$) covers 9,380 $\rm deg^2$ of the sky and overlaps with the western Galactic hemisphere on about 4,010 $\rm deg^2$  \citep{Strauss2002, AbazajianAdelman-McCarthyAgueros_2009ApJS..182..543A}. The stellar mass of the galaxies is estimated by spectral energy distribution (SED) fitting and has a typical uncertainty of 0.1 dex \citep{ChenKauffmannTremonti_2012MNRAS.421..314C}. The spectroscopic redshift ($z_{\rm spec}$) of the galaxy is estimated with an uncertainty of $\Delta z_{\rm spec}<10^{-4}$ \citep{Blanton2005}. The star-formation rate (SFR) of the galaxy is estimated by \citet{Brichmann2004}, which we take to model the XRB emission in Sect.~\ref{Sec_xrb}.
The galaxies hosting optical quasars identified with SDSS spectroscopy are not included in the SDSS MGS \citep{Strauss2002}. The galaxies hosting AGN in the SDSS MGS are identified as "AGN" or "composite" based on the BPT diagram, which we took to estimate the AGN emission, as described in Sect.~\ref{Sec_agn} \citep{BPT1981,Brichmann2004}.

\subsubsection{LS DR9 galaxy catalog}

The LS DR9 ($r_{\rm AB}<$ 23.4) covers 14,000 deg$^2$ of the extragalactic sky and overlaps with the western Galactic hemisphere on about 9,340 deg$^2$ \citep{Dey2019}. 
We use the galaxy catalog from \citet{Zou2019,Zou2022}\footnote{In this work, we use "LS DR9" to refer to the galaxy catalog from \citet{Zou2019,Zou2022}.}. 
This catalog uses three optical bands (g, r, z) from CTIO/DECam and two infrared bands (W1, W2) from WISE photometry. 
This catalog includes galaxies with $r_{\rm AB}<23$, $S/N>10$, and non-point-source-like morphology.
The photometric redshift ($z_{\rm phot}$) of the galaxy is estimated using a local linear regression algorithm \citep[LePhare,][]{ArnoutsCristianiMoscardini_1999MNRAS.310..540A, IlbertArnoutsMcCracken_2006A&A...457..841I}. 
The uncertainty of $z_{\rm phot}$ is $\Delta z_{\rm phot}=0.01$. 
The stellar mass ($M_*$) of galaxies is derived from the spectral energy distribution (SED) fitting using theoretical stellar population synthesis models from \citet{BruzualCharlot_2003MNRAS.344.1000B} at the photometric redshift. The uncertainty of $\log_{10}(M_*)$ is $\sim 0.2$ dex. 

The nearby galaxies with a large extent on the sky need special treatment in the DESI legacy survey as summarized in the Siena Galaxy Atlas (SGA) catalog \citep{MoustakasLangDey_2023}. 
We found that about 16,000 nearby galaxies in SGA are missing in the LS DR9 catalog. We added them to the catalog (see details in Appendix~\ref{Apd_catalog}).

\begin{table*}
\begin{center}
\caption{Description of the galaxy samples.}

\begin{tabular}{ccccccccccccccc}
\hline \hline 
\multicolumn{2}{c|}{$\log_{10}(M_*/M_\odot)$}&$M_{\rm *,med}/M_\odot$&\multicolumn{3}{|c}{$\log(M_{\rm 200m, CEN}/M_\odot)$}&\multicolumn{3}{|c|}{redshift}&\multicolumn{1}{|c|}{FULL$_{\rm spec}$}&\multicolumn{1}{|c|}{CEN}&\multicolumn{1}{|c}{SAT} \\
min&max&&16\%&med&84\%&min& max&med&$N_g$& $N_g$ &$N_g$\\
\hline
10.0&10.5&$1.8\times10^{10}$&11.5&11.75&12.1&0.01&0.06&0.05&11922&7956&3966 \\
10.5&11.0&$5.5\times10^{10}$ &11.8&12.1&12.6&0.02&0.10&0.08&45248&30825&14423 \\
11.0&11.25&$1.3\times10^{11}$&12.2&12.7&13.1&0.02&0.15&0.12&34046&26099&7947 \\
11.25&11.5&$2.2\times10^{11}$&12.6&13.1&13.6&0.03&0.19&0.15&24098&20342&3756 \\
\hline
\\
\end{tabular}

\begin{tabular}{ccccccccccc} 
\hline \hline 
\multicolumn{2}{c|}{$\log_{10}(M_{\rm 200m}/M_\odot)$}&$M_{\rm 200m,med}/M_\odot$&\multicolumn{3}{|c}{$\log(M_{\rm *,CEN_{halo}}/M_\odot)$}&\multicolumn{3}{|c|}{redshift}&\multicolumn{1}{|c}{CEN$_{\rm halo}$}&\multicolumn{1}{|c}{SAT$_{\rm halo}$}\\
min&max&&16\%&med&84\%&min& max&med&$N_g$&$N_g$\\
\hline
11.5&12.0&$5.5\times10^{11}$&9.7&10.2&10.6&0.01&0.08&0.06&25368&903 \\
12.0&12.5&$1.7\times10^{12}$&10.4&10.7&11.0&0.02&0.13&0.09&41301&4140 \\
12.5&13.0&$5.0\times10^{12}$&10.8&11.0&11.2&0.02&0.16&0.12&30740&9145 \\
13.0&13.5&$1.6\times10^{13}$&11.1&11.25&11.4&0.03&0.20&0.14&19889&9785 \\
13.5&14.0&$4.8\times10^{13}$&11.2&11.4&11.6&0.03&0.20&0.16&8214&12537 \\
\hline
\\
\end{tabular}

\begin{tabular}{cccccccccccc} 
\hline \hline 
\multicolumn{2}{c|}{$\log_{10}(M_*/M_\odot)$}&$M_{\rm *,med}/M_\odot$&\multicolumn{3}{|c|}{redshift}&\multicolumn{1}{|c|}{FULL$_{\rm phot}$}&\multicolumn{1}{|c}{Isolated}\\
min&max&&min& max&med&$N_g$& $N_g$ \\
\hline
9.5&10.0&$5.5\times10^{9}$&0.01&0.09&0.07&95698&24280 \\
10.0&10.5&$1.8\times10^{10}$&0.01&0.12&0.10&214286&50137 \\
10.5&11.0&$5.5\times10^{10}$ &0.02&0.17&0.14&415627&70348 \\
11.0&11.25&$1.3\times10^{11}$&0.02&0.33&0.28&627757&50766 \\
11.25&11.5&$2.2\times10^{11}$&0.03&0.40&0.35&324541&17983 \\
\hline
\end{tabular}

\tablefoot{For each table, from left to right, the columns are the stellar (halo) mass bin, median stellar (halo) mass, 16$^{\rm th}$ percentile, median, and 84$^{\rm th}$ percentile halo (stellar) mass of central galaxies, redshift bin, median redshift, and number of galaxies in the sample. 
Top: The FULL$_{\rm spec}$, CEN, and SAT galaxy samples built from SDSS MGS. FULL$_{\rm spec}$ contains all galaxies within the stellar mass bin and redshift bin. We split the FULL$_{\rm spec}$ sample into central (CEN) or satellite (SAT) galaxy samples according to the group finder algorithm \citep{Tinker2021}.
Middle: The CEN$_{\rm halo}$ and SAT$_{\rm halo}$ sample of central galaxies binned in halo mass, built from SDSS MGS and is based on the group finder algorithm \citep{Tinker2021}.
Bottom: The FULL$_{\rm phot}$ and isolated galaxy samples built from DESI LS DR9. The FULL$_{\rm phot}$ contains all galaxies within the stellar mass bin and redshift bin. We select the isolated sample with the algorithm defined in Appendix.~\ref{Sec_iso}.}
\label{Tab:gal:samples}
\end{center}
\end{table*}

\subsection{The galaxy samples}\label{Sec_sample_sum}

All galaxy samples in this work are built to be approximately volume-limited. 
We outline their key properties and what we use them for. Five galaxy samples were selected from the SDSS MGS spectroscopic galaxy catalog:
    \begin{itemize}
    \item{FULL$_{\rm spec}$} sample (Sect.~\ref{Sec_fullspec}): It includes all galaxies in the stellar mass range $10.0<\log(M_*)<11.5$ and spectroscopic redshift $0.01<z_{\rm spec}<0.19$ (top panel of Table~\ref{Tab:gal:samples}). It includes centrals and satellites. Therefore, we stack at the position of both centrals and satellites.
    We use this sample to help interpret the satellite boost bias (Sect.~\ref{Sec_profilefullsat}). We do not use this sample to derive the X-ray emission from the hot CGM.
    \item{CEN} sample (Sect.~\ref{Sec_cen}): It aims to include only central galaxies from the FULL$_{\rm spec}$ sample. The central galaxies are identified by the halo-based group finder with a misclassification possibility of about 1\% \citep{Tinker2021}. We use this sample to measure the X-ray surface brightness profiles of the hot CGM.
    \item{SAT} sample (Sect.~\ref{Sec_cen}): It includes only satellite galaxies from the FULL$_{\rm spec}$ sample. We use this sample to estimate the contamination from the misclassified central galaxies in the CEN sample. 
    \item{CEN$_{\rm halo}$} sample (Sect.~\ref{Sec_cen}): It aims to include only central galaxies selected in the halo mass range $\log M_{\rm 200m}=11.5-14.0$ (middle panel of Table~\ref{Tab:gal:samples}). The halo mass is estimated by the halo-based group finder algorithm \citep{Tinker2021}. We use this sample to measure the X-ray surface brightness profiles of the hot CGM at different halo masses.
    \item{SAT$_{\rm halo}$} sample (Sect.~\ref{Sec_cen}): It includes only satellite galaxies selected by the halo mass from the FULL$_{\rm spec}$ sample. We use it to estimate the contamination from the misclassified central galaxies for the CEN$_{\rm halo}$ sample. 
    \end{itemize}
Two galaxy samples are selected from DESI LS DR9 photometric galaxy catalog:
    \begin{itemize}
    \item{FULL$_{\rm phot}$} sample (Sect.~\ref{Sec_full}): It contains all galaxies in the stellar mass range $9.5<\log(M_*)<11.5$ and photometric redshift $0.01<z_{\rm phot}<0.40$ (bottom panel of Table~\ref{Tab:gal:samples}). The sample includes both centrals and satellites. Covering a larger area than SDSS, their stacked profiles will yield the highest signal-to-noise ratio (S/N). 
    \item{Isolated} sample (Appendix.~\ref{Sec_iso}): It includes galaxies selected from FULL$_{\rm phot}$ that live in under-dense environments. The sample has a high purity but a low completeness of isolated galaxies. It supplements the CEN samples and enables measurements of the total X-ray surface brightness profile at lower stellar mass. 
    \end{itemize}

\subsection{Building of the Full$_{\rm spec}$ sample from the SDSS DR7 and the Full$_{\rm phot}$ sample from the LS DR9 galaxy catalogs}\label{Sec_sdssls}

\subsubsection{Full$_{\rm spec}$ galaxy sample}\label{Sec_fullspec}
We selected galaxies with stellar mass $10.0<\log(M_*)<11.5$ from SDSS MGS because outside this range, the galaxy samples are too small, and identifying the central galaxy is hard. 
We split the galaxies into four stellar mass bins as listed in Table~\ref{Tab:gal:samples}. 
We name the $\log(M_*)=10.5-11.0$ bin as the "MW-mass" bin, the $\log(M_*)=11.0-11.25$ bin as the "M31-mass" bin, and the $\log(M_*)=11.25-11.5$ bin as the "2M31-mass" bin\footnote{The MW-mass, M31-mass, and 2M31-mass galaxies are defined only by the stellar mass, no extra selection is made on, for example, SFR or galaxy morphology}.
The maximum redshift is selected to ensure the sample is >95\% complete, namely being approximately volume-limited. 
Considering the fact that the hot CGM of very local galaxies may extend across a large area over the sky and be susceptible to inaccurate background subtraction, we leave them for future studies. We take minimum redshift $z_{\rm min}=0.01-0.03$ for galaxy samples at different $M_*$ bins. We obtain 115,314 galaxies in the FULL$_{\rm spec}$ sample.

\subsubsection{Full$_{\rm phot}$ galaxy sample}\label{Sec_full}

We select the galaxies with $9.5<\log(M_*)<11.5$ from the LS DR9 catalog. We split the galaxy sample into five stellar mass bins as listed in Table~\ref{Tab:gal:samples}. For the galaxy with both $z_{\rm spec}$ and $z_{\rm phot}$ measurements, we take $z_{\rm spec}$. 
Constrained by the PSF of eROSITA, we take different maximum redshifts ($z_{\rm 2\times PSF}$) for each mass bin (see Table~\ref{Tab:gal:samples}). Similar to FULL$_{\rm spec}$, we take minimum redshift $z_{\rm min}=0.01-0.03$ for galaxy samples at different $M_*$ bins. 
Finally, we obtained the 1,677,909 galaxies that constitute the FULL$_{\rm phot}$ sample. 

\subsection{Galaxy-halo connection}\label{Sec_galhalo}

Galaxies reside in various environments, ranging from voids to superclusters. To interpret the stacked measurement of the hot CGM without ambiguity, we need to relate one galaxy sample to its environment, that is, its host dark matter halo distribution \citep[see a review from][]{WechslerTinker_2018ARA&A..56..435W}, as discussed in \citet{Comparat2022}. 
Within a dark matter halo, the central galaxy, which is the most massive one, is thought to reflect the properties of the halo (i.e., the halo mass), and profoundly influences the hot CGM of the satellites; on the other hand, there is a loose relation between satellite galaxies and the halo properties.

Studies of galaxy clustering and galaxy-galaxy lensing constrained the galaxy-halo connection using SDSS galaxy survey at low redshift \citep{ZehaviZhengWeinberg_2011ApJ...736...59Z,ZuMandelbaum_2016MNRAS.457.4360Z}, 
the GAMA galaxy survey at slightly higher redshifts \citep{LinkeSimonSchneider_2022A&A...665A..38L}, and deeper spectroscopic surveys including COSMOS and CFHTLenS/VIPERS at even higher redshifts \citep{Leauthaud2012a,Coupon2015}. 
Overall, a double power-law with a 0.15 dex scatter can describe the stellar-to-halo relation for central galaxies well. There is no tight relation between the stellar mass of the satellite galaxy and its host halo mass.
The fraction of galaxies that are satellites of a more massive halo ($f_{\rm sat}$) decreases with increasing stellar mass; for the MW-mass galaxies, $f_{\rm sat}$ is about 30\%.

\subsection{Contamination by the satellite boost and X-ray sources in satellite galaxies}\label{Sec_proj}

\begin{figure}
        \centering
        \includegraphics[width=1.0\columnwidth]{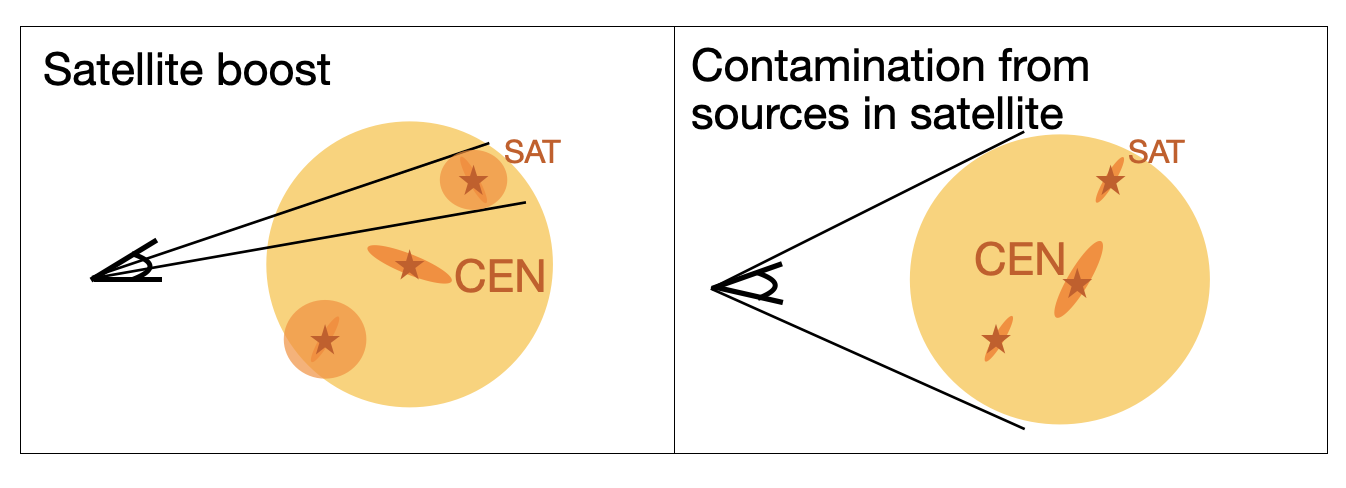}
        \includegraphics[width=0.97\columnwidth]{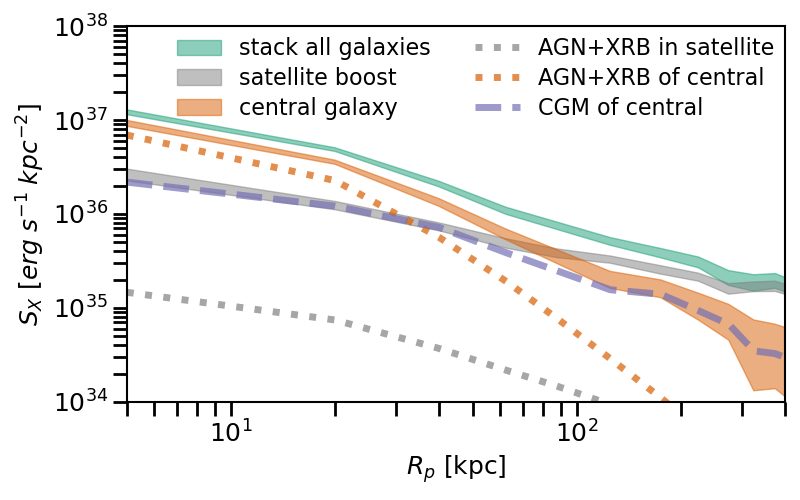}
        \caption{ Illustration of contamination to the stacking from the satellite boost bias and sources in the satellites (top). An illustration of the X-ray surface brightness profiles of different components is also shown in the bottom panel. The X-ray surface brightness profile of central galaxies that we are interested in (orange band) is obtained by subtracting the satellite boost (or misclassified central, gray band) from the stacking of all (or central) galaxies (green band). The hot CGM X-ray emission of the central galaxy (purple dashed line) is modeled by subtracting the X-ray emission from the X-ray sources in satellite galaxies (gray dotted line) and central galaxies (orange dotted line) from the X-ray surface brightness profile of central galaxies (orange band). 
        }
        \label{Fig_sketch}
\end{figure}

The existence of satellite galaxies influences the stacking in two ways (see top panel of Fig.~\ref{Fig_sketch} for an illustration): 
\begin{itemize}
    \item{Satellite boost}: Including satellite galaxies in the stacking sample would wrongly boost the X-ray emission. A satellite galaxy would be located in a more massive dark matter halo compared to a central galaxy with the same stellar mass. Therefore, the stacked X-ray emission around satellite galaxies is boosted by the hotter and brighter emission of the plasma from the more massive (parent) dark matter halo. This extra emission should not be attributed to the satellite galaxy. The bright X-ray emission from the nearby, more massive central galaxy flattens the X-ray profile of the satellite (see the filled curves in the bottom panel of Fig.~\ref{Fig_sketch}, more discussion in Sect.~\ref{Sec_profilesat}). 
    To avoid the satellite boost, we stack only central galaxies to measure the X-ray emission around galaxies. The misclassification of a satellite galaxy as a central galaxy causes a satellite boost in the stack, we model their contamination in Sect.~\ref{Sec_satmodel}. In the article we interchangeably use `misclassified central' (the cause) or `satellite boost' (the effect) to describe this feature.

\item{Contamination from sources in satellites}: Even when stacking around central galaxies, the X-ray emission from non-CGM X-ray sources (AGN or XRB) in satellite galaxies is unavoidably stacked and contaminates the X-ray emission from the hot CGM (see the dashed lines in the bottom panel of Fig.~\ref{Fig_sketch}). We will model the contamination by the X-ray sources in the satellite galaxies in Sect.~\ref{Sec_satsrc}.
\end{itemize}

In summary, the accurate selection of central galaxies and weeding out of the satellite contaminants represent a crucial element of our analysis. We discuss how this can be achieved for the spectroscopic samples at our disposal in the following sections. The discussion of the photometric sample is in Appendix.~\ref{Sec_iso}.

\subsection{Building of the CEN, SAT, CEN$_{\rm halo}$ and SAT$_{\rm halo}$ samples from SDSS MGS}\label{Sec_cen}

An unbiased selection of central galaxies requires accurate measurement of galaxy redshifts, the typical redshift uncertainty needs to be at least smaller than $10^{-4}$. 
For the SDSS MGS, it is possible to define a central galaxy sample, and different schemes exist to select central galaxies. A commonly used approach is the friends-of-friends (FoF) algorithm: galaxies are identified as a group according to the linking lengths, and within each group, the brightest galaxy is defined as the central galaxy \citep{CrookHuchraMartimbeau_2007ApJ...655..790C,RobothamNorbergDriver_2011MNRAS.416.2640R, TempelKipperTamm_2016AA...588A..14T}. 
An alternative method is to define the locally brightest galaxies as central galaxies and other galaxies falling within a certain distance around them are satellites \citep{Anderson2015, Comparat2022}.
More involved approaches include self-calibrated halo-based group finding \citep{YangMovandenBosch_2005MNRAS.356.1293Y, Tinker2022}, group finder with galaxy weighting \citep{AbdullahWilsonKlypin_2018ApJ...861...22A}, Bayesian group finder based on marked point processes \citep{TempelKruuseKipper_2018AA...618A..81T}. 

Here, the self-calibrated halo-based group finding algorithm is applied to the SDSS MGS to classify galaxies as centrals or satellites and estimate their halo mass \citep{Tinker2021, Tinker2022}. The halo-based group finding algorithm searches for galaxy groups within the same dark matter halos and identifies the central galaxy. 
The algorithm is calibrated by the mock galaxy distributions and is further self-calibrated using the observations, including the galaxy clustering measurements, lensing, and satellite luminosity measurements.
The algorithm assigns halos to galaxies by comparing the catalog to complementary data \citep{Tinker2021}. 
The algorithm's performances are excellent: the halo masses of either star-forming or quiescent central galaxies agree well with the halo masses estimated from weak-lensing observation. The clustering properties, halo occupation distribution (HOD), and $M_{\rm h}-M_*$ relation of star-forming or quiescent galaxies are also well reproduced \citep{Tinker2022}. The chance of misclassified central is about 1\%.

For each galaxy in the SDSS MGS, its probability as central or satellite ($P_{\rm sat}$, we use the recommended $P_{\rm sat}=0.5$ as a boundary), and its host halo mass is provided in \citet{Tinker2021}. 

\subsubsection{Central and satellite galaxy samples in stellar mass bins (CEN and SAT)}

The central galaxy sample (CEN) and satellite galaxy sample (SAT) selected from the SDSS MGS in stellar mass bins are listed in Table~\ref{Tab:gal:samples}. 
In total, we have 85,222 central galaxies and 30,092 satellite galaxies.

\subsubsection{Central and satellite galaxy samples in halo mass bins (CEN$_{\rm halo}$ and SAT$_{\rm halo}$)}

There is a large scatter in the stellar-to-halo-mass relation, which introduces the `inversion problem'\footnote{If sample A, selected based on stellar mass, has the same mean halo mass as sample B, selected based on halo mass, the galaxies in the two samples will be different (i.e., the stellar mass and SFR \citep[see Fig.2 in ][]{Moster2020}). This difference arises because the criteria used to select the galaxies are different. The inversion problem refers to the phenomenon where different selection criteria can lead to different sets of galaxies. Consequently, this can result in drawing different conclusions from samples A and B.}. Therefore, we also stack galaxies selected in halo mass.
With halo mass assigned to each galaxy in the SDSS MGS by the group finder algorithm, we can study how the $M_{\rm halo}$ ($M_{\rm 200m}$)\footnote{The $M_{\rm halo}$ provided in the catalog \citet{Tinker2021} is $M_{\rm 200m}$, the mass within the radius where the mean interior density is 200 times the background universe density.} is related to the X-ray emission of galaxies directly.
We define a halo mass-selected central galaxy sample (we name it as CEN$_{\rm halo}$ sample) by splitting it into five logarithmic bins in the range $\log M_{\rm 200m}=11.5-14.0$ of 0.5 dex each, see Table~\ref{Tab:gal:samples}. 
We set the maximum redshift for each CEN$_{\rm halo}$ bin at the redshift where its median stellar mass completeness is at 90\% for a magnitude cut $r_{AB}=17.77$. We have 125,512 central galaxies in the CEN$_{\rm halo}$ sample and 36,510 satellite galaxies in the SAT$_{\rm halo}$ sample. 

By selecting the CEN and CEN$_{\rm halo}$ samples respectively, we manage to account for the scatter in the SHMR. In Table~\ref{Tab:gal:samples}, we list the 16$^{\rm th}$ percentile, median, and 84$^{\rm th}$ percentile $M_{\rm 200m}$ ($M_*$) of each $M_*$ ($M_{\rm 200m}$) bin in the CEN (CEN$_{\rm halo}$) sample.
Although a stellar mass bin in the CEN sample and a halo mass bin in the CEN$_{\rm halo}$ sample may have similar median stellar mass or halo mass, the included galaxies are not the same. For example, the $\log(M_*)=10.5-11.0$ bin of CEN has a broader distribution of $M_{\rm 200m}\approx11.8-12.6$, compared to the $\log(M_{\rm 200m})=12.0-12.5$ bin of CEN$_{\rm halo}$.

\section{Models}\label{Sec_model}

Given the complexity of the galaxy-halo relation, the possible contaminations from satellite galaxies, and unresolved X-ray sources (XRB and AGN), a robust interpretation of the stacking results requires reliable models to control all of the above effects. 

In this section, we introduce a mock galaxy catalog built from simulations to help control the properties of galaxy samples in Sect.~\ref{Sec_mockcat}. The X-ray emission from unresolved AGNs and XRBs in the stacked galaxies and satellite galaxies is modeled in Sect.~\ref{Sec_agnxrb}. The contamination from the misclassified central galaxy in the group finder is modeled in Sect.~\ref{Sec_satmodel}.
We summarize how the models are used in Sect.~\ref{Sec_model_sum}.

\subsection{Mock galaxy catalog}\label{Sec_mockcat}

We built a mock catalog to help characterize and gain insights into the designed galaxy samples. 
We used the Uchuu simulation \citep{Ishiyama2021} and its \textsc{UniverseMachine} galaxy catalog \citep{BehrooziWechslerHearin_2019MNRAS.488.3143B}.
The Uchuu simulation is a suite of ultra-large cosmological N-body simulations to construct the halo and sub-halo merger trees in a box of $2.0 \rm h^{-1} Gpc$ side-length, with mass resolution of $3.27\times 10^{8} \rm h^{-1} M_\odot $, using the cosmological parameters obtained by Planck \citep{Planck2020,Ishiyama2021}. 
The galaxy model aptly reproduces the stellar mass function and the separation between red-sequence and blue-cloud galaxies. 
We created a full-sky light cone following the method from \citet{Comparat2020}. 

We empirically calibrated the relation between stellar mass and K-band absolute magnitude (and its scatter) as a function of redshift using observations from SDSS, KIDS+VIKING, GAMA (for redshift smaller than 0.75) and COSMOS (for higher redshifts) \citep{Ilbert2013, AlmeidaAndersonArgudo-Fernandez_2023,KuijkenHeymansDvornik_2019A&A...625A...2K, Driver2022}. 
With these relations, we assign an absolute K-band magnitude to each simulated galaxy using the stellar mass prediction from \textsc{UniverseMachine}.
Then, we search the observed data sets for the nearest neighbor to each simulated galaxy in redshift and K-band. 
The set of observed broadband magnitudes (particularly the r-band) obtained with the match is assigned to the simulated galaxy. 
We find that the galaxy painting process is faithful to the observed galaxy population up to redshift $\sim$0.5\footnote{Beyond the redshift of 0.5, systematic effects due to the observed samples used (less accurate photometric redshifts in KIDS, small area subtended by COSMOS) lessen the quality of the mock catalog. However, we do not use the mock catalog in this redshift regime.}. 
The light cone well reproduces the observed K-band and r-band luminosity functions from GAMA (or SDSS). 
We can thus apply the magnitude limit used in this work to the light cone. In the light cone, central (as the brightest galaxy) and satellite galaxies are defined accurately.

We built a mock SDSS MGS catalog from the light cone to z=0.45. We apply the magnitude cut $r_{AB}<17.77$ and add a Gaussian distributed $z_{\rm error}=10^{-4}$ and $M_{\rm error}=0.01\ \rm dex$ to the accurate ones from the light cone. We use the mock SDSS MGS catalog to evaluate the satellite galaxy distribution. 

\subsection{Models for AGN and XRB emission}\label{Sec_agnxrb}

The PSF and sensitivity of eROSITA prevent us from resolving all point sources at the redshift of the galaxies we stack. 
Therefore, to infer what fraction of the measured X-ray emission comes from the hot CGM, in this section we opt for an empirical model to predict the XRB X-ray emission based on the stellar mass and SFR of galaxies (Sect.~\ref{Sec_xrb}). 
For the AGN, we opt for an observation-based approach (Sect.~\ref{Sec_agn} and Appendix.~\ref{Apd_agn}). The X-ray emission of XRB in the satellite galaxies is modeled using the mock catalog (Sect.~\ref{Sec_satsrc}). 
With the predicted luminosity of the XRB and AGN, we normalize the PSF to obtain their X-ray surface brightness profiles\footnote{We notice XRB are located around the galaxy disk and its surface brightness profile may not be PSF-like. To test if it has an impact, we do the following. We assume XRB distribute as a S$\Acute{\rm e}$rsic profile: $S_{\rm X}=S_{\rm X,0} \exp [-(r/r_{50})^{1/n}]$, where $r$ is the distance to galaxy center, $r_{50}$ is the half-light radius of the galaxy and $n$ is the S$\Acute{\rm e}$rsic index. We conservatively take $n=4$ and $r_{50}=10\,\rm kpc$ \citep[see][Fig. 11]{Shen2003}. We convolve the S$\Acute{\rm e}$rsic profile with the PSF and find the difference to the PSF to be negligible ($<5\%$). Therefore, we use only the PSF to model the XRB surface brightness profile.}. 

\subsubsection{Model for XRB}\label{Sec_xrb}

The X-ray luminosity of XRB in the 2-10 keV band ($L_{\rm 2-10\,keV}$) scales with the SFR and $M_*$ of host galaxies \citep{Aird2017,Lehmer2019}. For the CEN and CEN$_{\rm halo}$ samples, we predict the $L_{\rm 2-10keV}$ and estimate its uncertainty using the model in \citet{Aird2017} (see Eq.5 of \citet{Aird2017}). We also use the empirical model compiled by \citet{Lehmer2019} to estimate the XRB luminosity and obtain consistent results. The selection of the XRB model does not affect our conclusions. 
We convert $L_{\rm 2-10keV}$ to the X-ray luminosity in 0.5-2 keV ($L_{\rm XRB}$) by assuming an absorbed power law with a photon index of 1.8, and with column density fixed at $N_H=2.3\times10^{20}\rm cm^{-2}$, the mean $N_H$ of the stacked area. 

\subsubsection{Model for AGN}\label{Sec_agn}

To model the AGN emission for the CEN and CEN$_{\rm halo}$ samples, we consider the galaxies identified as "AGN" or "composite" using the BPT diagram applied to the SDSS MGS central galaxies and name these galaxies as CENAGN sample \citep{BPT1981,Brichmann2004}. 
Depending on the stellar mass, $f_{\rm AGN}=$8-18\% of the galaxies in the CEN sample were selected as CENAGN (see Table~\ref{Tab_cenagn}). 
We stacked the galaxies in the CENAGN sample, masking all detected X-ray sources to calculate the maximum X-ray emission from unresolved AGN and XRB as described in Appendix.~\ref{Apd_agn}. We estimated the AGN emission ($L_{\rm AGN}$) as the residual emission after subtracting the XRB emission. Under the assumption of no other galaxies hosting AGN, the lower limit of the average AGN contamination in the CEN sample is $L_{\rm AGN}\times f_{\rm AGN}$. To be conservative, we take the upper limit of AGN contamination as $2\times L_{\rm AGN}\times f_{\rm AGN}$; this accounts for (obscured) X-ray AGN not being identified as AGN or composite in the SDSS spectra. We repeated the procedure for the CEN$_{\rm halo}$ sample to estimate the AGN contamination there (see the discussion in Appendix.~\ref{Apd_agn}).
We find that the unresolved AGN contamination is comparable to the XRB emission at the stellar mass (or halo mass) ranges we focus on. Our AGN model is consistent with the empirical AGN model derived in \citet{Comparat2019, Comparat2023}.

\subsubsection{XRBs in satellite galaxies} \label{Sec_satsrc}

The AGN and XRBs in the satellites would contaminate the hot CGM emission attributed to the central galaxy. 
We took the mock SDSS MGS catalog to evaluate the satellite galaxy distribution. 
For each stacked galaxy in the CEN sample, we selected the galaxy with similar stellar mass (difference smaller than 0.1 dex) and redshift (difference smaller than 0.01) in the mock catalog. We counted the number of satellite galaxies and the projected distances of the satellites to the mock central galaxy. 
The AGN population in satellite galaxies is unclear; there could be 0-20\% satellite galaxies hosting AGN \citep{Comparat2023}. We assume the AGN contamination in the satellite galaxies is negligible after applying masks.
We estimate the XRB emission of the satellites according to properties ($M_*$ and SFR) from the mock catalog, as in Sect.~\ref{Sec_xrb}. We sum and average the XRB emission in satellites as a function of distance to the central galaxy. 

We find the total X-ray luminosity of unresolved XRBs in satellites is about 8\% (19\%, 37\%, 66\%) of the XRB in the stacked central galaxies for the four $M_*$ bins in the CEN sample, respectively. Therefore, the model of unresolved point sources in satellites is necessary, especially for massive galaxies. We repeat the same procedure above for the CEN$_{\rm halo}$ sample.

\subsection{Model for misclassified central galaxies}
\label{Sec_satmodel}
The satellite galaxies might be misclassified as central in the group finder, with a chance of <1\% \citep{Tinker2021}. 
In the case of stellar mass selection (CEN), since satellites are hosted by more massive (brighter in X-ray) haloes, the contamination can not be ignored.
Here, we consider the 1\% satellite galaxies in the CEN (or CEN$_{\rm halo}$) sample and model their contamination. We stack galaxies and measure the X-ray emission in the SAT (or SAT$_{\rm halo}$) sample ($S_{\rm X,SAT}$). We rescale the X-ray profiles of CEN ($(S_{\rm X,CEN}$) as $(S_{\rm X,CEN}-1\% S_{\rm X,SAT})/99\%$. The rescaled X-ray luminosity of CEN drops by about 5\%.

\subsection{Application of the models}\label{Sec_model_sum}

The models were applied to the CEN and CEN$_{\rm halo}$ samples as follows: 
\begin{itemize}
    \item{Modelling the misclassified central}. We follow the procedure in Sect.~\ref{Sec_satmodel} to rescale the measured X-ray surface brightness profile of CEN and CEN$_{\rm halo}$ samples. This step corrects for the contamination from the 1\% misclassified central (namely satellite) galaxies. 
    \item{Modelling the unresolved AGN and XRB in central galaxies}. We assume the mean X-ray profile of unresolved point sources to be PSF-like. We estimate the X-ray luminosity of unresolved AGN in the CEN and CEN$_{\rm halo}$ samples following the procedure in Sect.~\ref{Sec_agn}. We estimate the X-ray luminosity of XRB in the stacked central galaxies with the empirical model described in Sect.~\ref{Sec_xrb}. We sum up the X-ray luminosity of unresolved AGN and XRB to normalize a PSF profile accordingly. 
    \item{Modelling the unresolved XRB in satellite galaxies}. We consider that the satellite galaxies located within the halo of stacked galaxies contaminate the hot CGM. We estimate the XRB luminosity of the satellite galaxies following the procedure in Sect.~\ref{Sec_satsrc}. The X-ray profile is predicted using the satellite distribution from the mock catalog.
    \end{itemize}

\noindent
From the stacked surface brightness profile ($S_{\rm X, stack}$) with X-ray point sources masked (see Sect.~\ref{Sec_method}) and misclassified centrals corrected (see Point 1 above), we obtain the surface brightness of the CGM by subtracting the modeled contamination ($S_{\rm X, AGN+XRB+SAT}$, Points 2 and 3 above): $S_{\rm X, CGM}=S_{\rm X, stack}-S_{\rm X, AGN+XRB+SAT}$. 

\section{X-ray surface brightness profile around galaxies from the FULL and SAT samples}\label{Sec_profilefullsat}

In this section, we present the measured X-ray surface brightness profiles of FULL$_{\rm phot}$ (or FULL$_{\rm spec}$) samples (Sect.~\ref{Sec_profilefull}). We quantify the satellite boost bias in Sect.~\ref{Sec_profilesat} and apply it to model the misclassified centrals as described in Sect.~\ref{Sec_satmodel}.

\subsection{X-ray surface brightness profile of FULL$_{\rm phot}$ sample} \label{Sec_profilefull}
\begin{figure}[h]
    \centering
    \includegraphics[width=1.0\columnwidth]{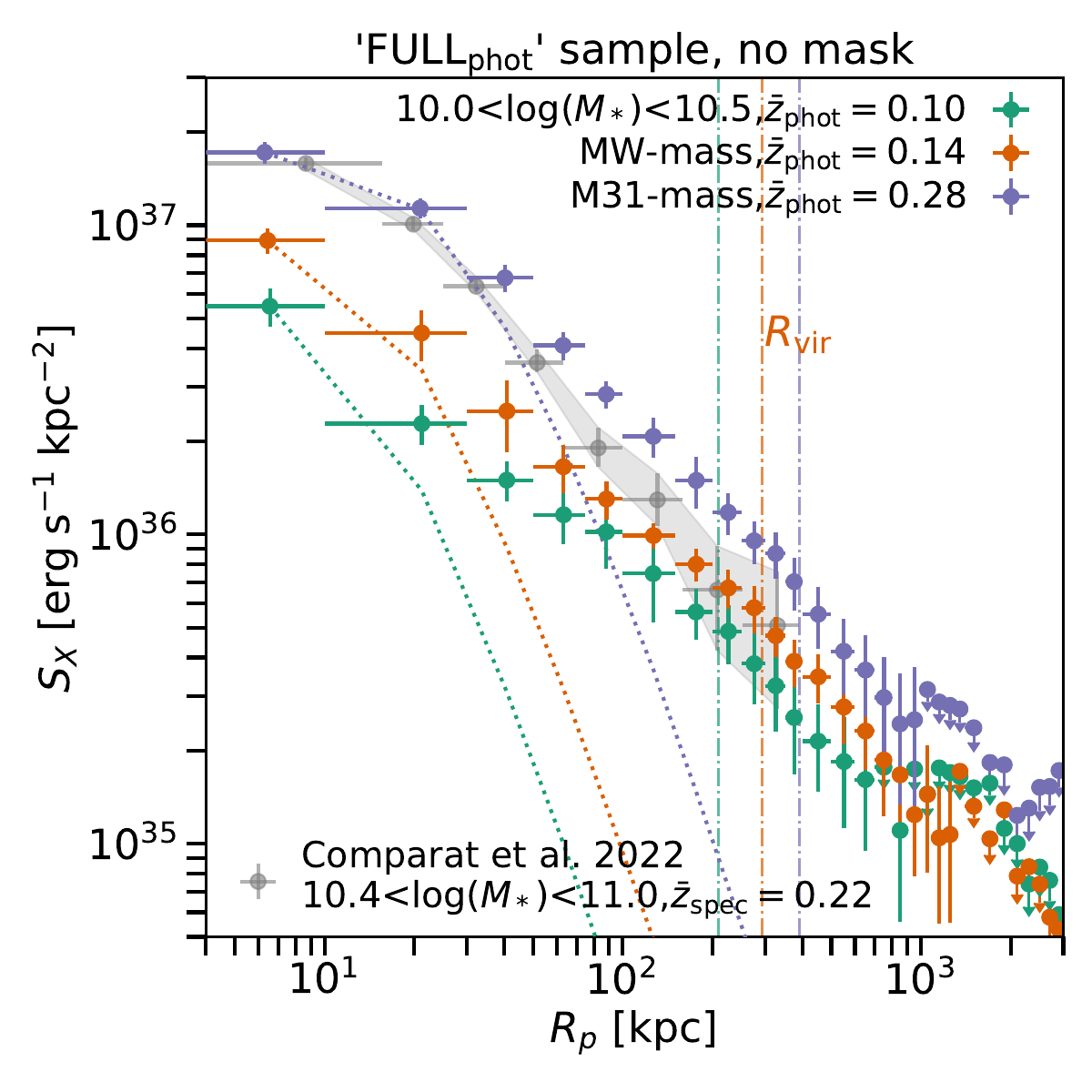}
    \caption{X-ray surface brightness profiles of the FULL$_{\rm phot}$ sample after subtracting background, without masking sources. The results are divided into galaxy mass bins $\log(M_*)=10.0-10.5$, 10.5-11.0 (MW-mass), and 11.0-11.025 (M31-mass) as labeled. The dotted line is the PSF of eROSITA, and the vertical dash-dotted line denotes the average virial radius of stacked galaxies. The X-ray emission comprises all effects mentioned in Sect.~\ref{Sec_proj}, limiting its interpretability regarding CGM emission.
    The X-ray surface brightness profile measured by stacking 16,142 galaxies with $\log(M_*)=10.4-11.0$ and $z_{\rm spec}=0.05-0.3$ is plotted in gray \citep{Comparat2022}. The galaxy sample in \citet{Comparat2022} contains a complete population of AGN (while our FULL$_{\rm phot}$ does not contain sources that appear point-like in the legacy survey photometry) and has a broader PSF due to its higher mean redshift. So we expect the previous measurement to agree with the FULL MW-mass at large separation but to be brighter in the center. 
}
        \label{Fig_profile_full_nomask}
\end{figure}

Stacking the FULL$_{\rm spec}$ or FULL$_{\rm phot}$ sample without masking any detected X-ray sources introduces the least uncertainty and bias (from X-ray data reduction or galaxy sample compiling) to the results; the measurement is reproducible. 
The surface brightness profiles of the FULL$_{\rm phot}$ samples in $\log(M_*)=10.0-10.5$, MW-mass and M31-mass (after subtracting the background) are shown in Fig.~\ref{Fig_profile_full_nomask}.

We notice the significant improvement of the statistics by comparing the X-ray surface brightness profile to the one measured by stacking 16,142 galaxies with $\log(M_*)=10.4-11.0$ and $z_{\rm spec}=0.05-0.3$ selected from the GAMA catalog using the eFEDS X-ray data \citep{Comparat2022}.
The X-ray surface brightness profiles obtained here have a high  S/N. 
These high-significance measurements benefit from a large number of sources available within the large area considered in this work and covered by the eROSITA observations. 

Due to the satellite boost bias we described in Sect.~\ref{Sec_proj}, the satellite galaxies located in more massive dark matter halo strongly influence the profiles shown in Fig.~\ref{Fig_profile_full_nomask}, resulting in the stacked X-ray emission extending to about 1-2 Mpc (about 5-10 times of $R_{\rm vir}$ of galaxies).
The X-ray sources contributing to these stacks include AGN, XRB, and hot gas emission: intra-cluster medium (ICM), intra-group medium (IGrM), and CGM\footnote{Hereafter in our work, we use CGM to denote the medium within the virial radius of the galaxy, regardless if the galaxy is in group or cluster.} from central and satellite galaxies in different massive halos. 
The interpretation of these measurements remains complex, and future simulation and modeling efforts will enable an unambiguous interpretation of all the signals present in these measurements. 
In Sect.~\ref{Sec_profilesat}, we briefly discuss the different X-ray surface brightness profiles around central and satellite galaxies. However, we leave the elaboration of an accurate satellite boost model that will enable the complete extraction of the information in these stacks with high S/N values for future studies (Shreeram et al. in prep). 

\subsection{Untangle the X-ray emission from satellite and central galaxies in FULL$_{\rm spec}$ sample}\label{Sec_profilesat}

\begin{figure}[h]
    \centering
        \includegraphics[width=1.0\columnwidth]{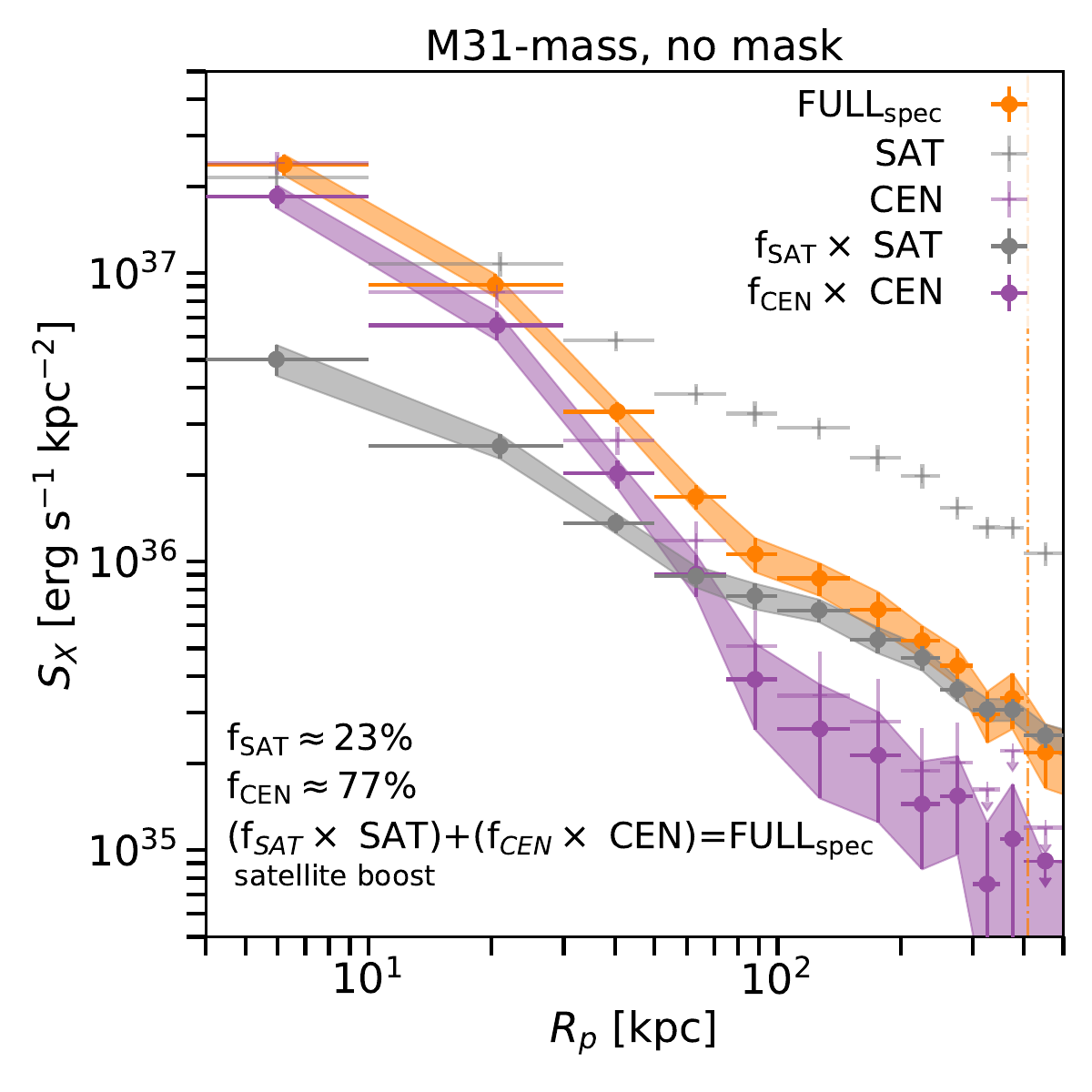}
    \caption{Observed mean X-ray surface brightness profile of galaxies in M31-mass bin of FULL$_{\rm spec}$ sample and the profiles after splitting FULL$_{\rm spec}$ sample to central (CEN) or satellite (SAT) galaxies. We scale the observed surface brightness profiles of CEN and SAT according to the number ratio of galaxies, namely, $f_{\rm SAT}=N_{\rm g,SAT}/N_{\rm g,FULL_{spec}}$ and $f_{\rm CEN}=N_{\rm g,CEN}/N_{\rm g,FULL_{spec}}$. The vertical dash-dotted line denotes the average virial radius of M31-mass galaxies in the FULL$_{\rm spec}$ sample. }
        \label{Fig_profile_censat}
\end{figure}

The satellite boost biases the results of the stacking experiment. In this section we use the central/satellite galaxy classification of the SDSS spectroscopic sample and stack the X-ray emission around galaxies in CEN and SAT samples to study the difference and to untangle the X-ray surface brightness profile of central galaxies from the stacking of FULL$_{\rm spec}$. 

The comparison of the X-ray surface brightness profile of FULL$_{\rm spec}$, CEN, and SAT sample is plotted in Fig.~\ref{Fig_profile_censat}, taking the M31-mass bin as an example. 
As expected, galaxies with the same $M_*$ but classified as satellites or centrals have different X-ray surface brightness profiles. The satellite galaxy, residing in a more massive dark matter halo, has its X-ray emission boosted by the emission from virialized gas of the larger halo they reside in, from nearby centrals and other satellite galaxies. Consequently, the X-ray surface brightness profiles of the satellite galaxies appear to be more extended and brighter than those of the central.  

We rescale the profile of central or satellite galaxies by their number ratio. Namely, we multiply the mean profile of central or satellite galaxies by $N_{CEN}/N_g$ or $N_{SAT}/N_g$, respectively. These rescaled profiles of centrals and satellites, compared to the FULL$_{\rm spec}$ sample, are plotted in Fig.~\ref{Fig_profile_censat}. We find that, beyond 100 kpc, the emission from satellite galaxies dominates over the central galaxy and determines the profile shape of the FULL$_{\rm spec}$ sample, as we had argued. This proves the necessity of selecting central galaxies to study how a galaxy modulates the properties of its CGM. On the other hand, we can use the rescaled profiles to untangle the emission in FULL$_{\rm spec}$ sample and model the influence of the misclassified central galaxy, as we discussed in Sect.~\ref{Sec_model}. Finally, we go on to focus on the CEN and CEN$_{\rm halo}$ samples where the satellite boost bias that hampers the interpretation of the full stacks is minimized.

\section{The circumgalactic medium of central galaxies selected in stellar mass and halo mass} \label{Sec_profilegal}

In this section, we mask all identified X-ray point sources so that the emission from the hot gas component becomes more discernible.
The observed surface brightness profiles of the CEN and CEN$_{\rm halo}$ samples are presented in Sect.~\ref{Sec_profileisocen}.
The X-ray emission from the hot CGM emission is modeled (following the steps in Sect.~\ref{Sec_model_sum}) and presented in Sect.~\ref{Sec_profileCGM}.

\subsection{The X-ray surface brightness profiles }\label{Sec_profileisocen}

\begin{figure*}[h!tb]
    \centering
    \includegraphics[width=1.0\columnwidth]{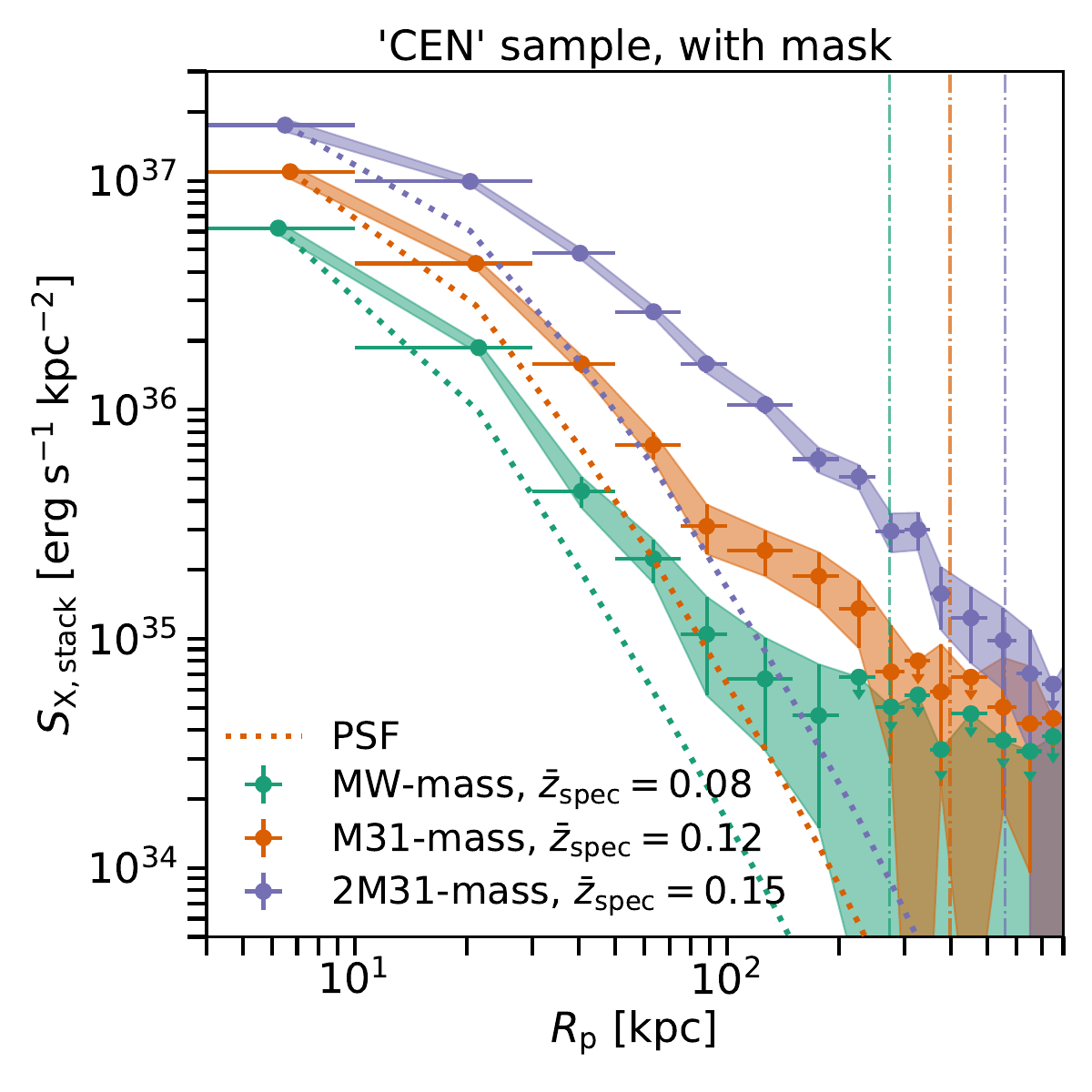}
    \includegraphics[width=1.0\columnwidth]{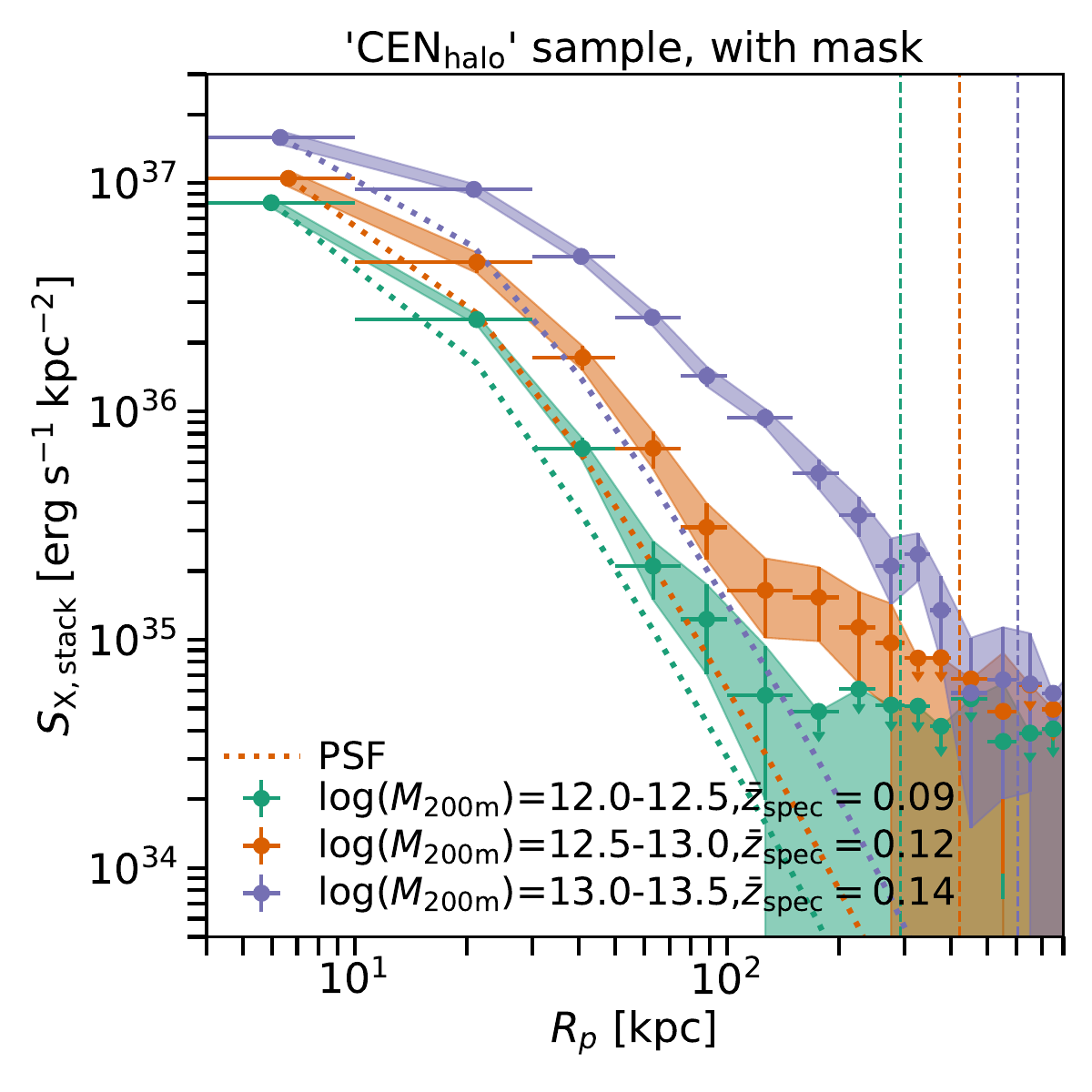}
    \caption{Mean X-ray surface brightness profiles of central galaxies (with a magnitude limit of $r_{\rm AB}<17.77$) after subtracting background and masking X-ray point sources, for galaxies with $\log(M_*)=10.5-11.0$ (MW-mass), 11.0-11.25 (M31-mass), and 11.25-11.5 (2M31-mass) shown on the left. The X-ray emission comprises unresolved point sources and hot gas. $\Bar{z}$ is the median redshift of galaxies in each mass bin. The dotted lines are the PSF of eROSITA. The vertical dash-dotted lines are the virial radius at different stellar mass bins correspondingly. Right panel shows the mean X-ray surface brightness profiles of central galaxies with $\log(M_{\rm 200m})=12.0-13.5$ after subtracting background and masking X-ray point sources.
    }
        \label{Fig_profile_ms}
\end{figure*}

The X-ray surface brightness profiles up to $R_{\rm vir}$
of the CEN sample in MW-mass, M31-mass, and 2M31-mass are plotted in the left panel of Fig.~\ref{Fig_profile_ms}. 
The X-ray emission detected here is the sum of unresolved point sources and hot gas emission from the stacked central galaxies (plus their satellite galaxies) within $R_{\rm vir}$. We find that the X-ray surface brightness increases with stellar mass. The extended nature of the emission can be visually identified by comparing the X-ray surface brightness profile to the PSF of eROSITA (dotted line).
Flatter profiles than the PSF are observed beyond 30 kpc for MW-mass and more massive galaxies (left panel of Fig.~\ref{Fig_profile_ms}). The profile becomes more extended with increasing stellar mass. 
For the MW-mass, M31-mass, and 2M31-mass galaxies in the CEN sample, the X-ray emission is detected within the virial radius of galaxies with $S/N \approx 3.1,4.7,8.3$.

The X-ray surface brightness profiles of galaxies with $\log(M_{\rm 200m})=12.0-13.5$ are plotted in the right panel of Fig.~\ref{Fig_profile_ms}. The X-ray surface brightness increases with $M_{\rm 200m}$. Extended emission is detected around central galaxies in halos with $\log(M_{\rm 200m})>12.5$ with $S/N \approx 2.8, 5.3, 6.9$ for the three bins, respectively.

For the first time, we measure the extended X-ray emission up to the virial radius around the MW-mass and M31-mass central galaxies with a high S/N. 

\subsection{The hot CGM emission in MW-mass or more massive galaxies} \label{Sec_profileCGM}

\begin{figure*}[h!tb]
    \centering
    \includegraphics[width=0.48\linewidth]{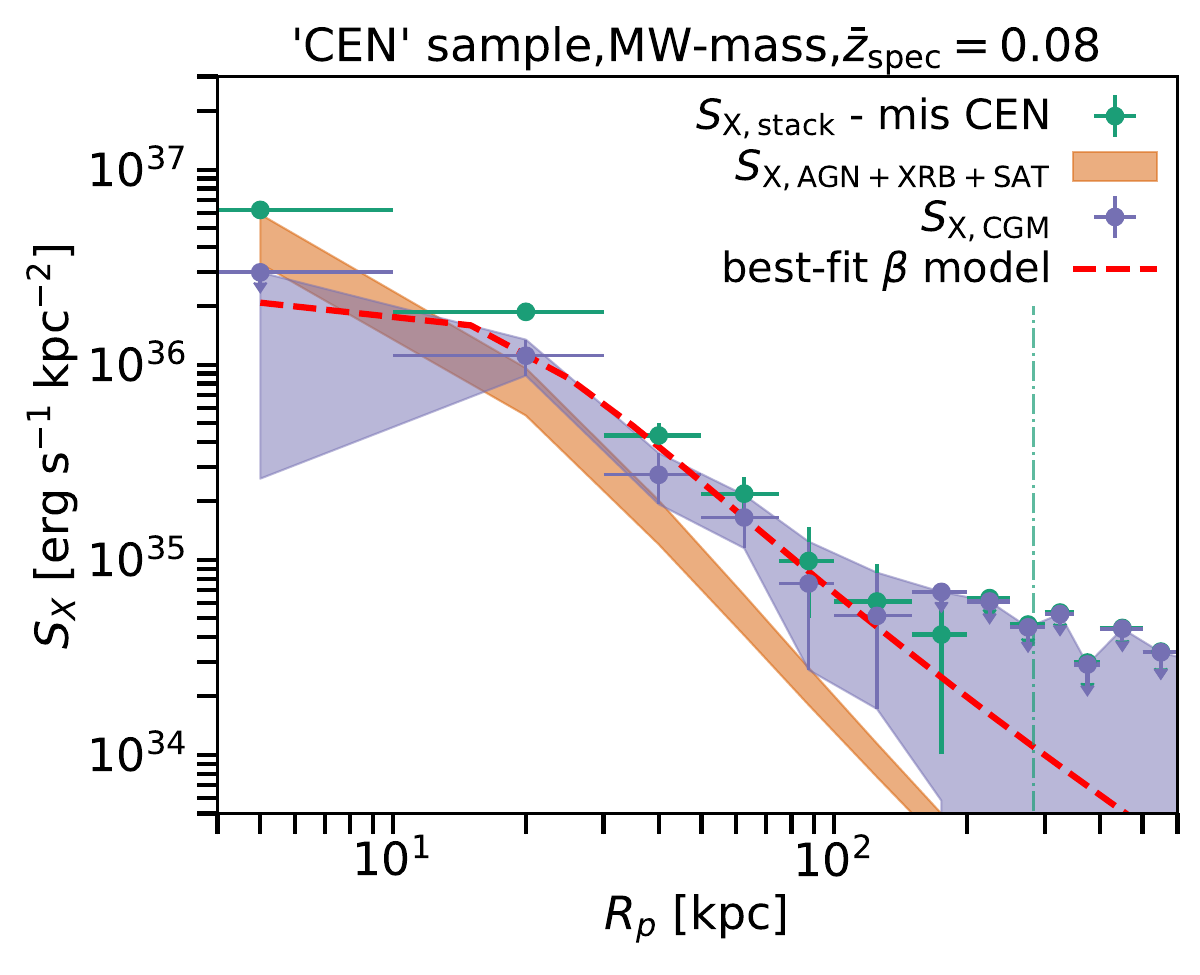}
    \includegraphics[width=0.48\linewidth]{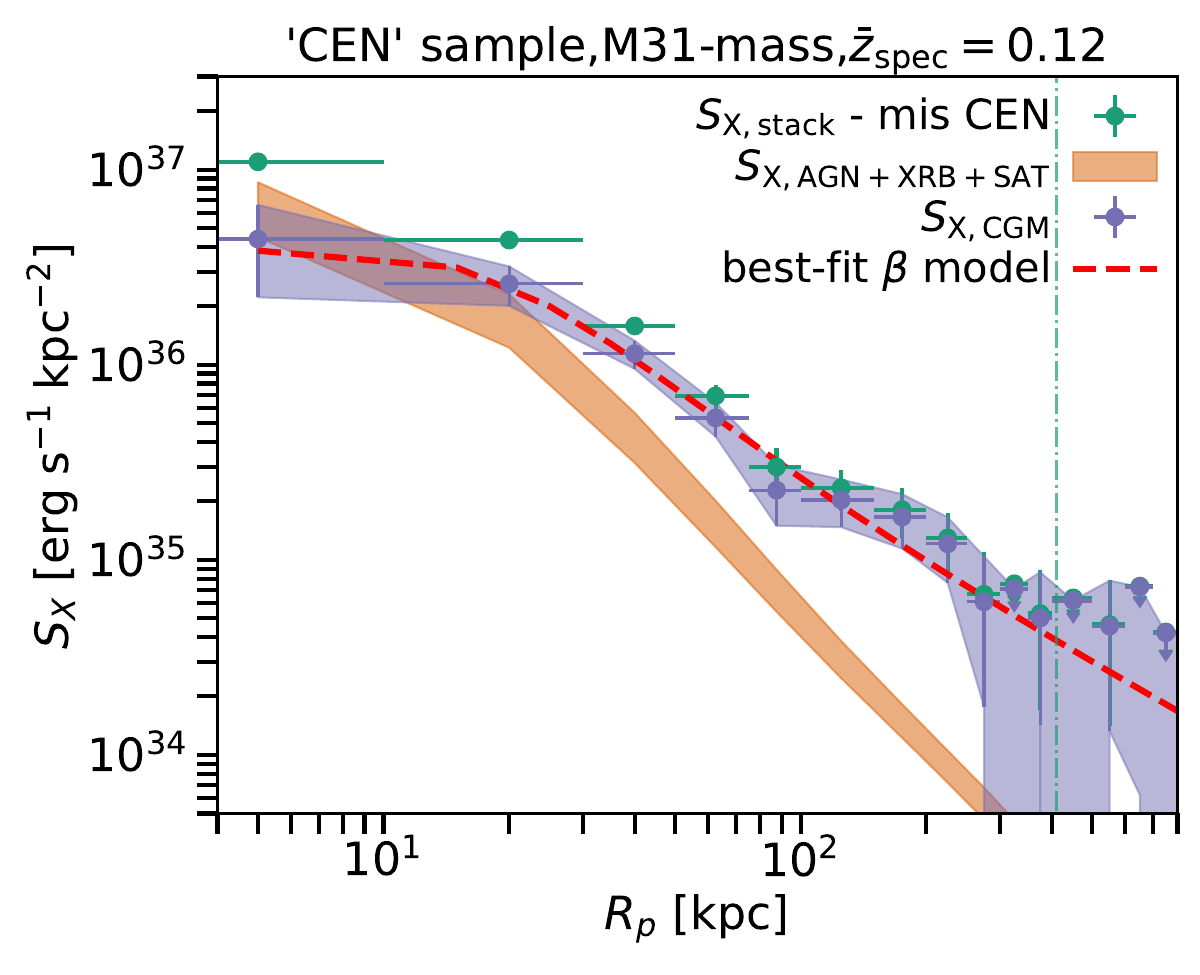}\\
    \includegraphics[width=0.8\linewidth]{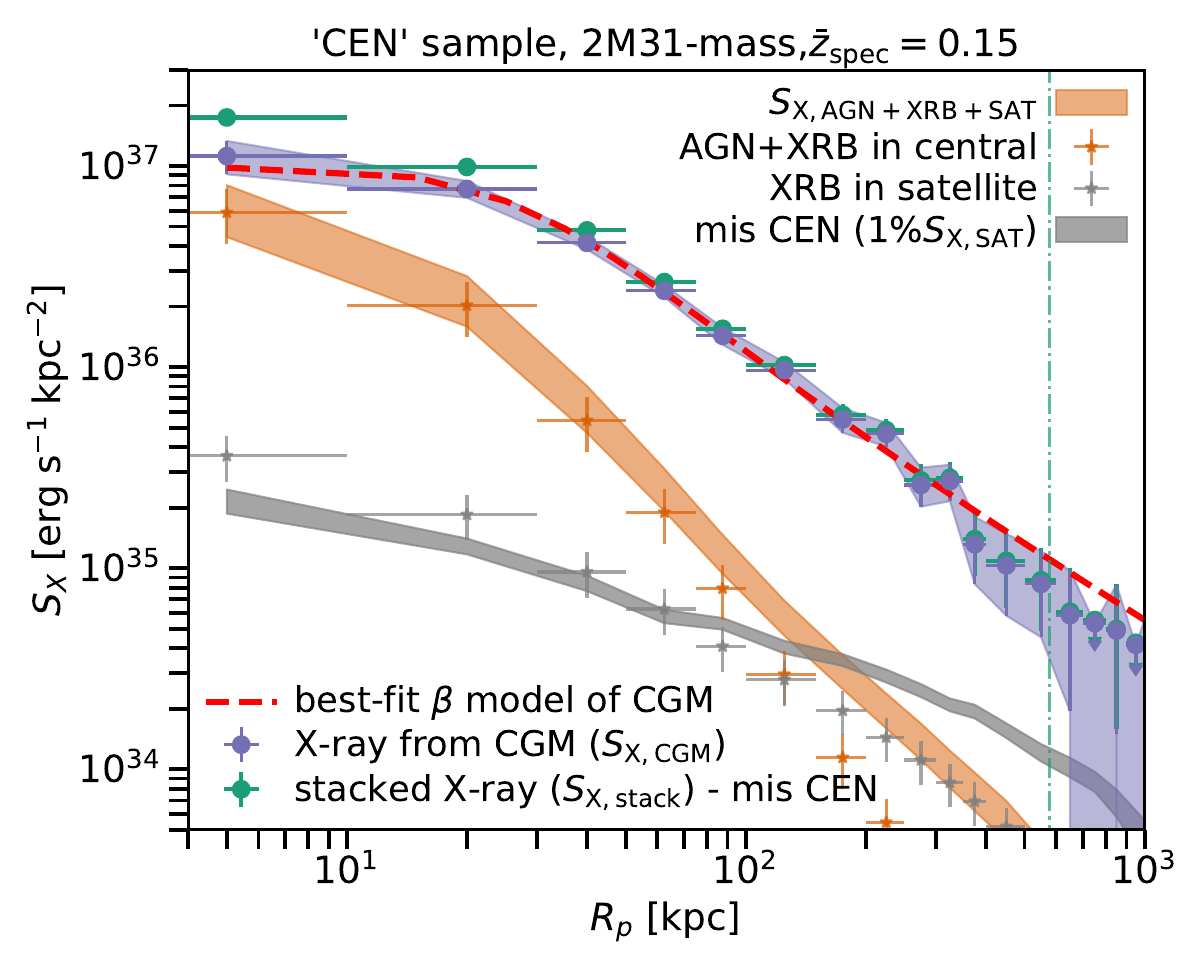}
    \caption{Mean X-ray surface brightness profiles of central galaxies selected from SDSS in different $M_*$ bins, after masking X-ray sources and correction of misclassified centrals (green points, $S_{\rm X,stack}$), the total X-ray surface brightness of unresolved AGN and XRB in the stacked galaxies and satellites (orange band), and the X-ray surface brightness of the CGM (purple band, $S_{\rm X,CGM}$) modeled by subtracting the unresolved AGN and XRB from the $S_{\rm X,stack}$ (See Sect.~\ref{Sec_model_sum} for details). The red dashed line is the best-fit of the CGM profile with a $\beta$ model. The vertical dash-dotted line is the virial radius, $R_{\rm vir}$. The top two panels are for MW-mass and M31-mass galaxies, and the bottom panel is for 2M31-mass galaxies. In the bottom panel, we further denote the X-ray surface brightness profiles of modeled unresolved AGN and XRB in the central galaxies (orange cross), XRB in the satellite galaxies (gray cross), and X-ray surface brightness profile of misclassified satellite (gray band).}
        \label{Fig_profile_cgm}
\end{figure*}

\begin{figure}[h!tb]
    \centering
    \includegraphics[width=1\columnwidth]{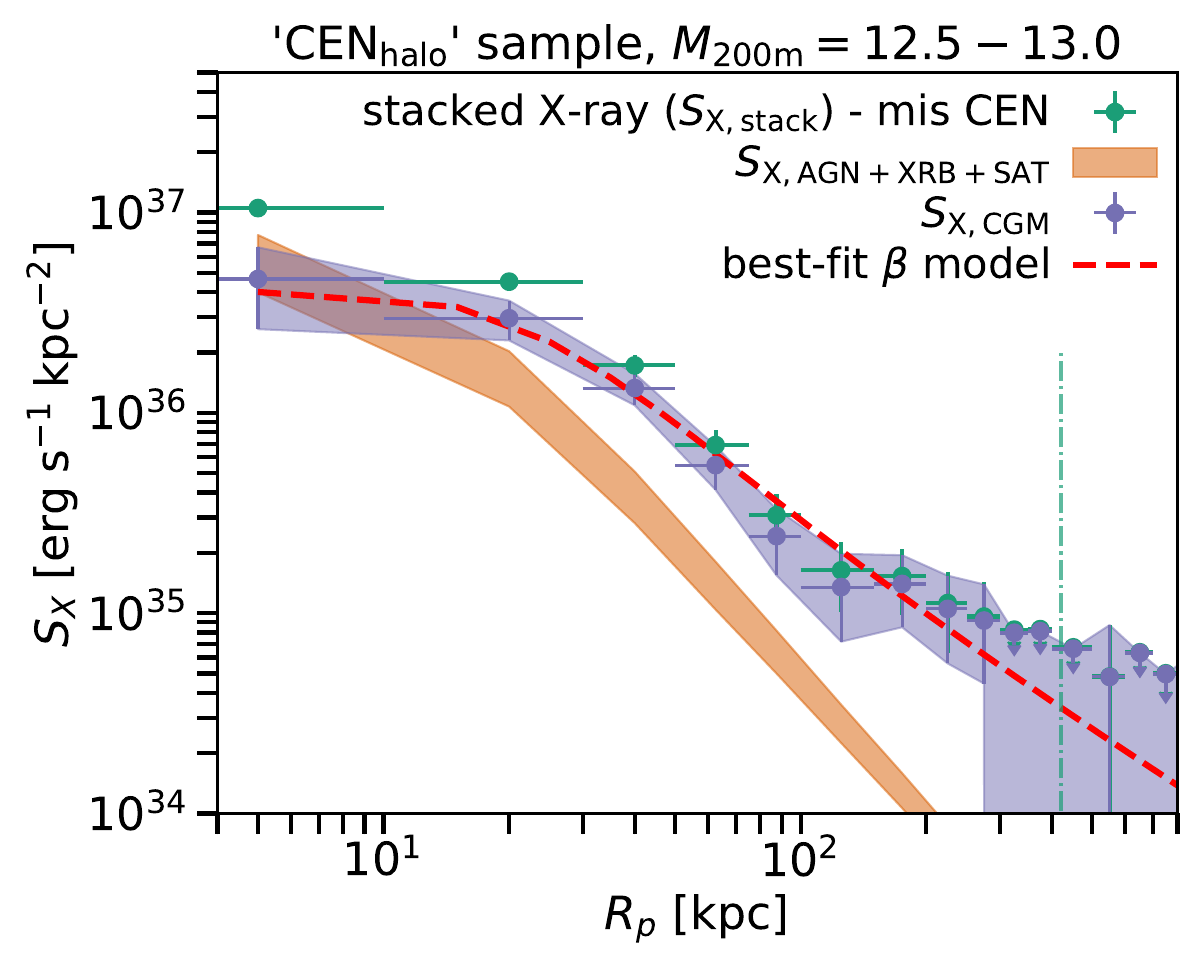}
    \includegraphics[width=1\columnwidth]{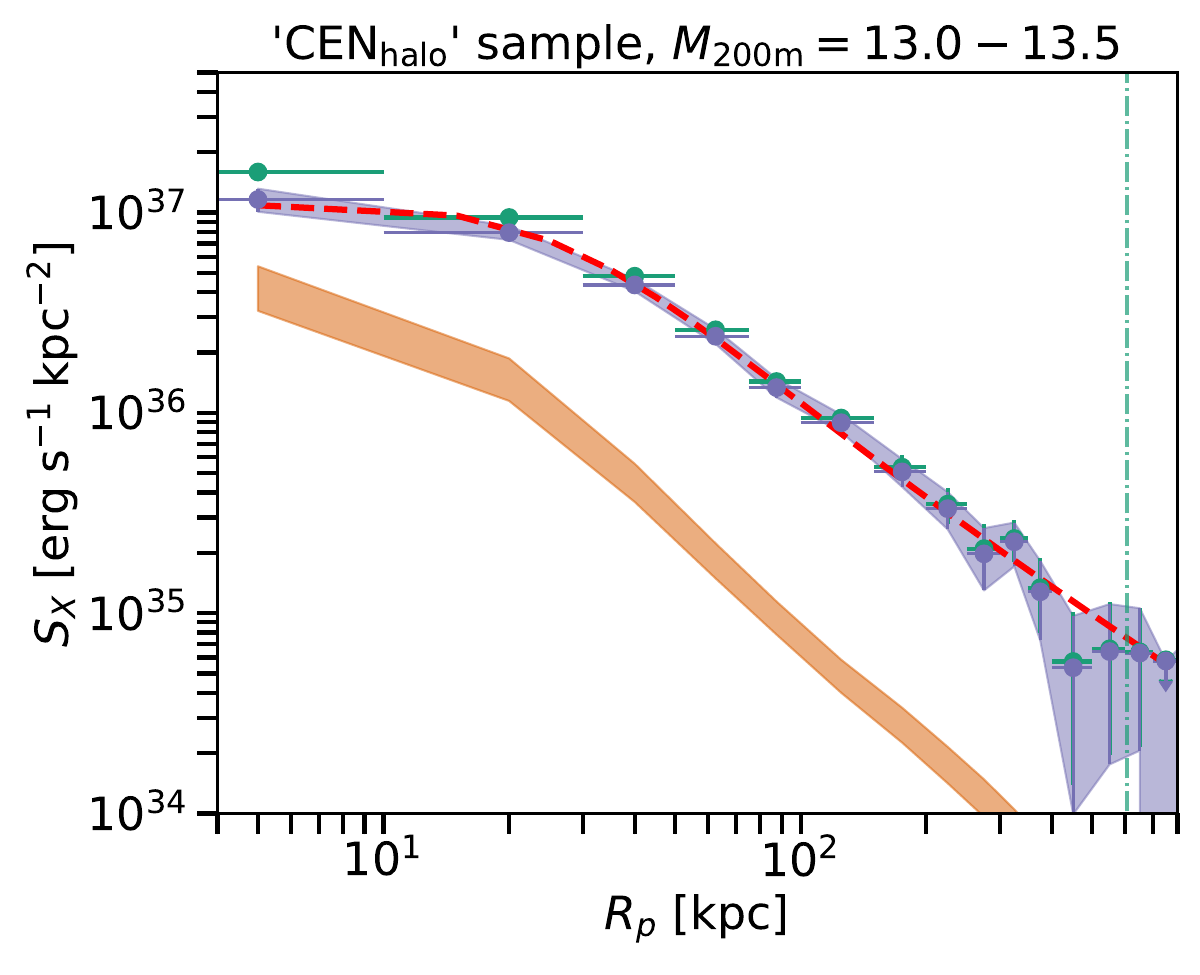}
    \includegraphics[width=1\columnwidth]{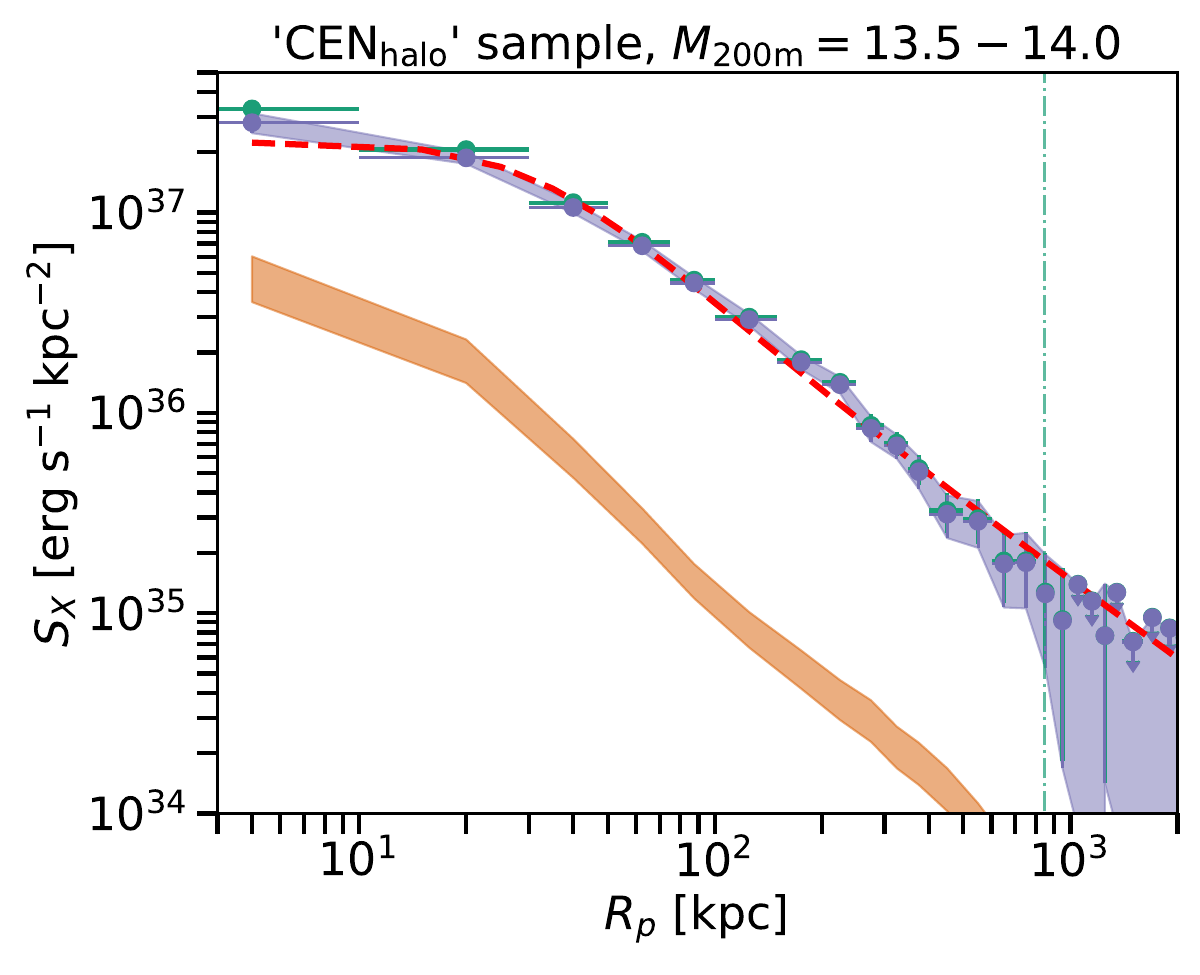}
    \caption{Same as Fig.~\ref{Fig_profile_cgm}, but for galaxies selected in halo mass bins. From top to bottom, the X-ray surface brightness profiles for $M_{\rm 200m}=12.5-13.0$, 13.0-13.5, 13.5-14.0.}
        \label{Fig_profile_halo_cgm}
\end{figure}

\begingroup
\renewcommand{\arraystretch}{1.5} 
\begin{table}[h!tb]
\centering
\caption{Best-fit parameters of $\beta$ model for the X-ray surface brightness profiles of CGM for the CEN and CEN$_{\rm halo}$ samples. }
\label{tab:betamodel}
\begin{tabular}{p{20mm}cccccc}
\hline \hline
&MW-mass&M31-mass&2M31-mass\\
\hline
$\beta$ & $0.43_{-0.06}^{+0.10}$&$0.37_{-0.02}^{+0.04}$&$0.37_{-0.01}^{+0.02}$ \\
$\log(S_{\rm X,0})$ \newline $[\rm erg\,s^{-1}\,kpc^{-2}]$&$36.7_{-0.4}^{+1.4}$&$37.1_{-0.4}^{+1.5}$&$37.3_{-0.1}^{+0.1}$\\
$r_{\rm c} [\rm kpc]$&$6_{-5}^{+9}$&$4_{-3}^{+7}$&$10_{-3}^{+3}$\\
\hline
$\chi^2$&4.0&3.2&6.6\\
d.o.f.&5&8&10\\
$\chi^2$/d.o.f.&0.8&0.4&0.7\\
\hline
\\
\end{tabular}

\begin{tabular}{p{20mm}cccccc}
\hline \hline
&&$\log(M_{\rm 200m})$& \\
&12.5-13.0&13.0-13.5&13.5-14.0\\
\hline
$\beta$    & $0.39_{-0.03}^{+0.06}$&$0.39_{-0.02}^{+0.02}$&$0.38_{-0.01}^{+0.01}$ \\
$\log(S_{\rm X,0})$ \newline $[\rm erg\,s^{-1}\,kpc^{-2}]$ &$37.2_{-0.3}^{+1.3}$&$37.3_{-0.1}^{+0.1}$&$37.6_{-0.1}^{+0.1}$\\
$r_{\rm c} [\rm kpc]$&$5_{-4}^{+8}$&$10_{-3}^{+3}$&$15_{-3}^{+3}$\\
\hline
$\chi^2$&4.5&4.4&16.0\\
d.o.f.&9&11&13\\
$\chi^2$/d.o.f.&0.5&0.4&1.2\\
\hline
\end{tabular}

\tablefoot{The uncertainty is of 1$\sigma$. We use the X-ray surface brightness profiles within $R_{\rm vir}$ for the fits.}
\end{table}
\endgroup

The hot CGM profiles of MW-mass, M31-mass, and 2M31-mass central galaxies are presented in Fig.~\ref{Fig_profile_cgm}. The modeled X-ray emission from AGN, XRB, and misclassified central galaxies (plotted in orange) contributes to most of the emission within 20 kpc and is less extended than the detected X-ray emission (green dots).
The X-ray emission from the hot CGM (purple band) is detected within $R_{\rm vir}$ with $S/N\approx 1.7, 3.9, 7.7$ for MW-mass, M31-mass and 2M31-mass galaxies.

We fit the $\beta$ model (Eq.~\ref{eq:Sbeta}) to the hot CGM X-ray profile to obtain an analytical description of the results. 
The fit results are reported in Table~\ref{tab:betamodel}.
We get $\beta \approx 0.43$ for MW-mass galaxies and $\beta \approx 0.37$ for M31-mass and 2M31-mass galaxies. $\log(S_{\rm X,0})$ increases with $M_*$, while $r_{\rm c}$ is poorly constrained, limited by the uncertainty of modeled AGN, XRB, and misclassified central galaxies emission.

The results for CEN$_{\rm halo}$ sample are presented in Fig.~\ref{Fig_profile_halo_cgm}. The X-ray emission from the hot CGM is detected within $R_{\rm vir}$ with $S/N\approx 2.2, 4.9, 6.6$ for the halo mass bins $\log(M_{\rm 200m})=12.5-13.0$, $13.0-13.5$, and $13.5-14.0$ respectively. The fit results of the $\beta$ model are listed in Table~\ref{tab:betamodel}. We get $\beta \approx 0.4$.


Prior to the publication of this work, the value of $\beta$ had been measured in several nearby massive galaxies ($\log(M_*)\approx 11.5$), within $0.5 R_{\rm vir}$ with $\beta\approx0.397\pm0.009$, which is consistent with our measurement\footnote{Besides the $\beta$ model, the exponential model is proposed to describe the \emph{inner} hot CGM `corona' of the Milky Wan and nearby galaxies, $S_{\rm X, exp}=S_{\rm X, 0} e^{-r/r_{\rm s}}$, where $r_{\rm s}$ is the scale radius with a value of some kiloparsecs \citep{Yao2009,Locatelli2023}. Limited by the PSF of eROSITA, we can not resolve the sharp corona component.} \citep{Li2017,Li2018}. 
The hot CGM density profile beyond $0.5 R_{\rm vir}$ is not well studied yet, as a result, the baryon budget stored in the hot CGM is poorly constrained by the X-ray observations \citep{Li2018,Das2020,DasChiangMathur_2023ApJ...951..125D}.
With our newly measured hot CGM density profile within $R_{\rm vir}$, we describe how we estimated the baryon mass and fraction in Sect.~\ref{Sec_baryon}.

\section{The baryon budget}
\label{Sec_baryon}

\begin{figure*}[h!tb]
    \centering
    \includegraphics[width=1.0\columnwidth]{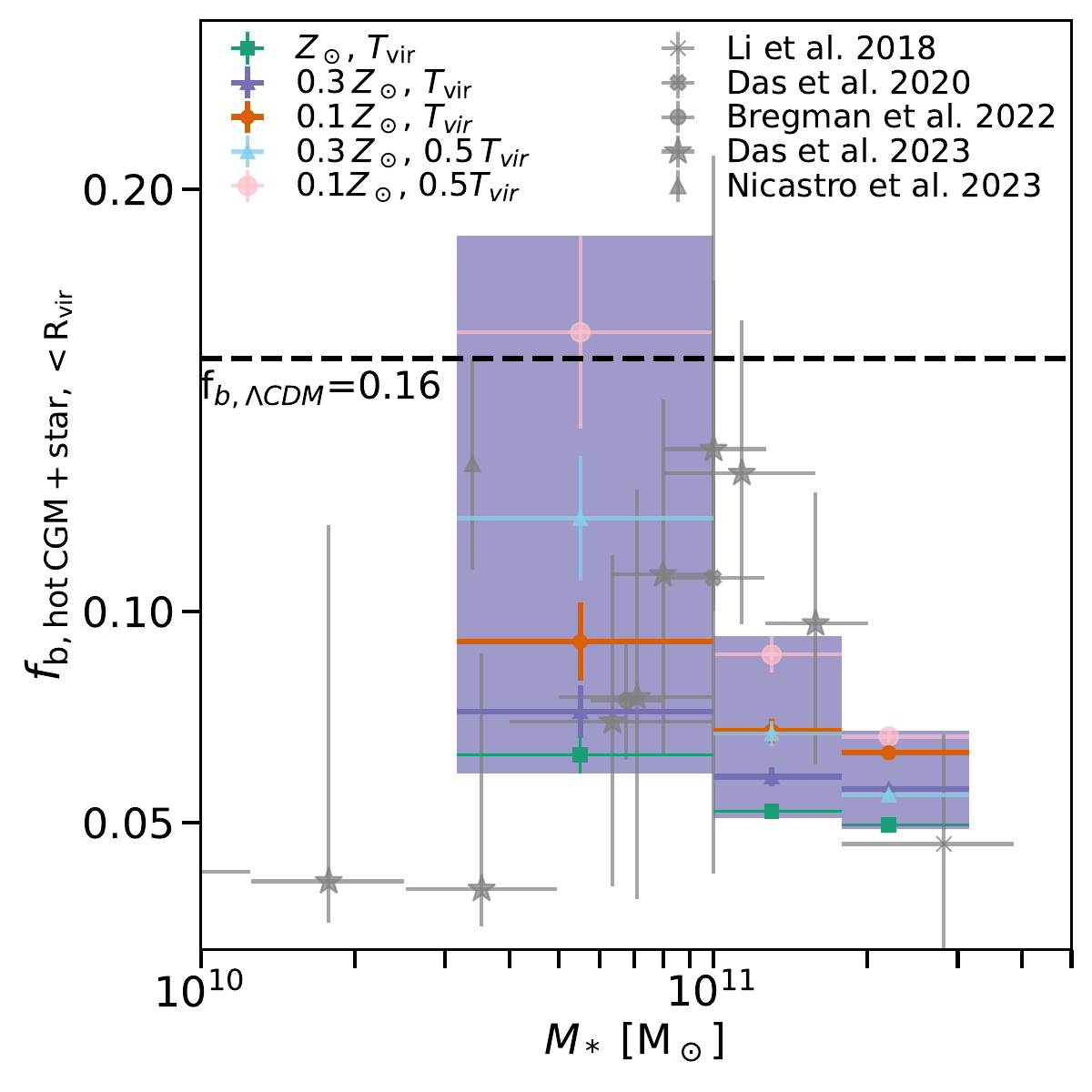}
    \includegraphics[width=1.0\columnwidth]{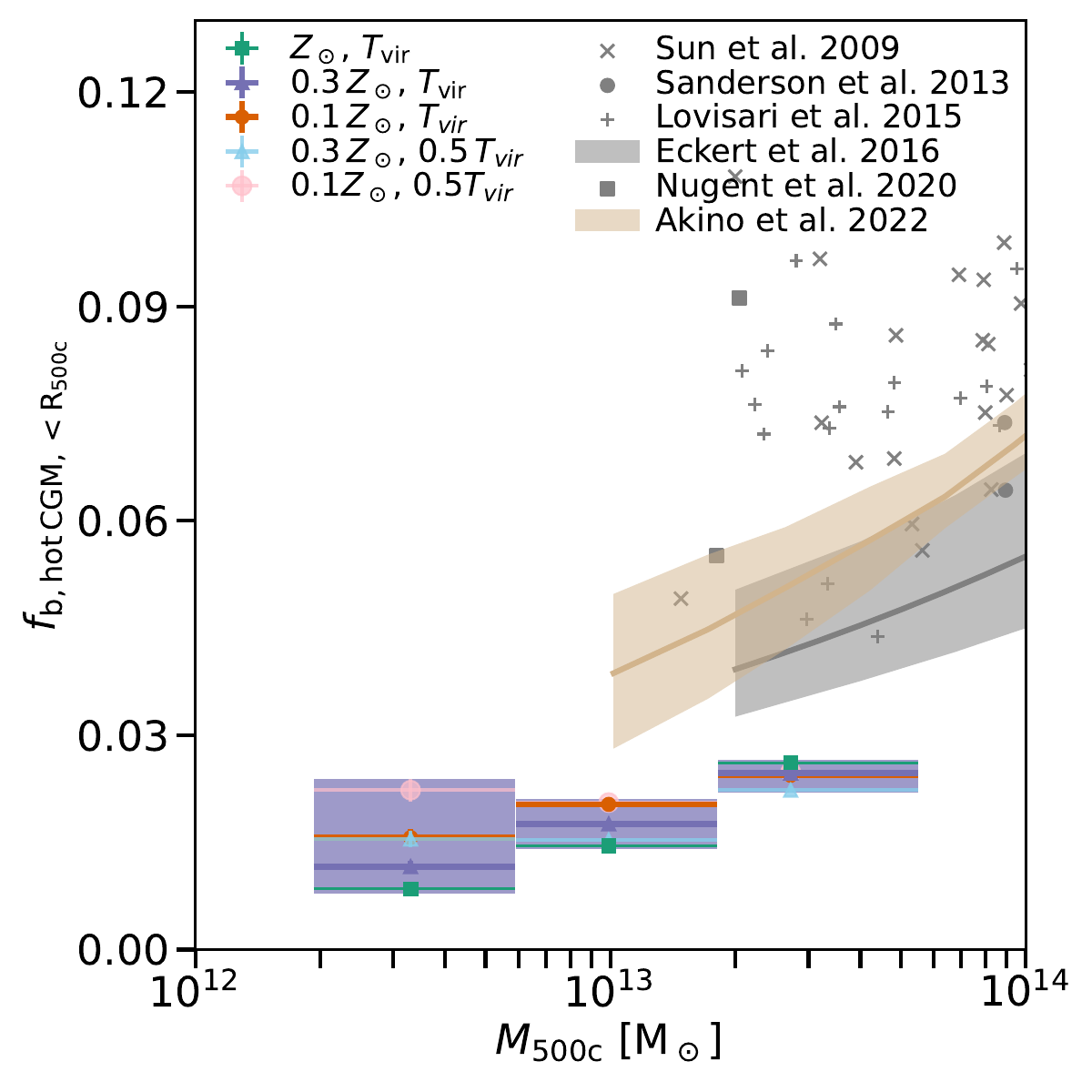}
    \caption{Baryon fraction as a function of $M_*$ given on the left. The baryon fraction is the sum of the stellar and the hot CGM within $R_{\rm vir}$ divided by the virial mass ($f_{\rm b, hot\,CGM+star,<R_{vir}}=(M_{\rm hot\,CGM,<R_{vir}}+M_*)/M_{\rm vir}$). We assume that the metallicity abundance of the hot CGM is $0.1Z_\odot$, $0.3Z_\odot$ and $Z_\odot$, and the temperature of the hot CGM is at virial temperature ($T_{\rm vir}$) or half ($0.5T_{\rm vir}$). The baryon fractions predicted with different metallicity abundances and temperatures are summarised in purple square. The dashed line ($f_{\rm b}=0.16$) is the cosmological baryon fraction \citep{Planck2020}. The gray points are taken from tSZ measurements \citep{Bregman2022, DasChiangMathur_2023ApJ...951..125D}, and X-ray measurements \citep{Li2018,Das2020,Nicastro2023}. Right panel shows the baryon fraction as a function of $M_{\rm 500c}$ under assumptions of metallicity abundances and temperatures. The baryon fraction is the hot CGM within $R_{\rm 500c}$ divided by the $M_{\rm 500c}$ ($f_{\rm b, hot\,CGM,<R_{500c}}=(M_{\rm hot\,CGM,<R_{500c}})/M_{\rm 500c}$).The gray points are gas fractions estimated from detected clusters and groups in X-ray \citep{Sun2009,Sanderson2013,Lovisari2015,Eckert2016,Nugent2020,Akino2022}.}
        \label{Fig_fbaryon}
\end{figure*}

\begin{figure}[h!tb]
    \centering
    \includegraphics[width=1.0\columnwidth]{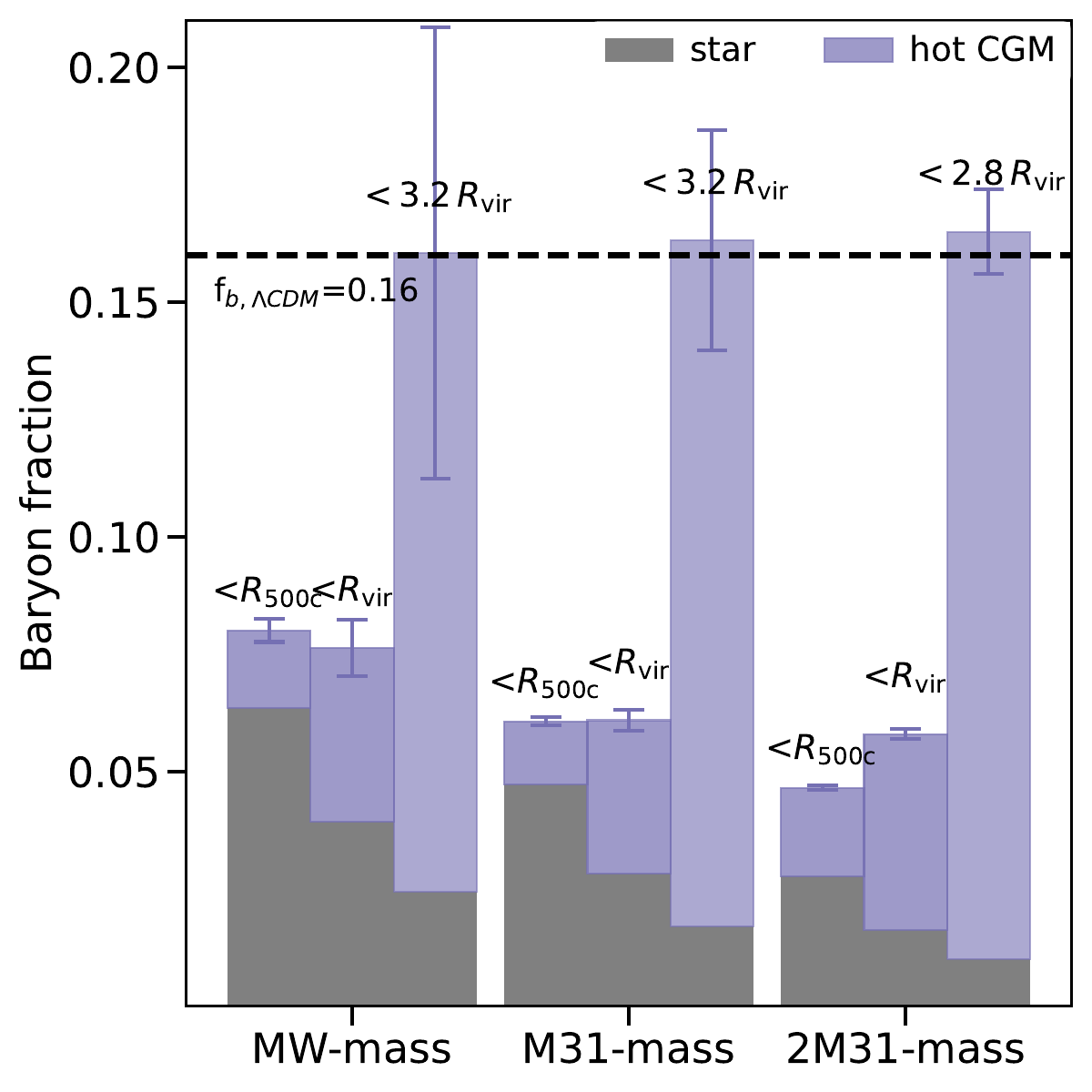}
    \caption{Baryon fraction within $R_{\rm 500c}$, $R_{\rm vir}$ and the radius within which the baryon fraction exceeds $16\%$ ($R_{\rm c}$), for MW-mass, M31-mass and 2M31-mass galaxies. We take the baryon fraction derived from the assumption of $T_{\rm vir}$ and $0.3Z_\odot$. The error bar is the uncertainty propagated from the $\beta$ model.}
        \label{Fig_fbaryon_hist}
\end{figure}

Based on the derived parameters of $\beta$ models (in Sect.~\ref{Sec_profileCGM}), we estimated the baryon mass and baryon fraction within $R_{\rm vir}$ of central galaxies. 
The density profile of the hot CGM in the $\beta$ model is then: 
\begin{equation}
    n(r)=n_{0}\left[1+\left(\frac{r}{r_{\rm c}}\right)^2\right]^{(-\frac{3}{2}\beta)}\ [\rm cm^{-3}],
\end{equation}
where $n_{0}$ is the gas density at the galaxy center and is expressed as

\begin{equation}
    n_0=\sqrt{\frac{S_{X,0}\Gamma(3\beta)}{\sqrt{\pi}\epsilon_{\rm X} \frac{n_{\rm e}}{n_{\rm H}}r_{\rm c}\Gamma(3\beta-1/2)} } [\rm cm^{-3}],
\end{equation}

where $n_{\rm e}$ and $n_{\rm H}$ are the number densities of electrons and hydrogens, $\Gamma$ is gamma function \citep{Ge2016}. We take the APEC model to estimate $\epsilon_{\rm X}(T, Z)$, which is the normalized X-ray flux of thermal plasma with temperature $T$ and metal abundance $Z$ \citep{Smith2001}. We assume the temperature of the hot CGM is at the virial temperature $T_{\rm vir}$, which can be derived from the $M_{\rm vir}$ and $R_{\rm vir}$ \footnote{For reference, the average $M_{\rm vir}$ of MW-mass, M31-mass, and 2M31-mass galaxies are taken as $1.3\times10^{12}M_\odot$, $4.2\times10^{12}M_\odot$, and $1.2\times10^{13}M_\odot$. The $T_{\rm vir}$ are taken as $0.22\,\rm keV$, $0.5\,\rm keV$, and $1.0\,\rm keV$. The median $M_{\rm vir}$ of the three bins in $\log(M_{\rm 200m})=12.5-14.0$ of CEN$_{\rm halo}$ sample are $5.0\times10^{12}M_\odot$, $1.6\times10^{13}M_\odot$, and $4.3\times10^{13}M_\odot$. The $T_{\rm vir}$ are taken as $0.57\,\rm keV$, $1.20\,\rm keV$, and $2.65\,\rm keV$. }, or half virial Temperature (to be conservative).
The X-ray emissivity of the hot CGM at $T_{\rm vir}<1 \rm keV$ depends strongly on the metallicity abundance, which is not well studied for the MW-mass galaxies yet. We take values of $0.1Z_\odot$, $0.3Z_\odot$ and $Z_\odot$ to estimate $\epsilon_{\rm X}(T, Z)$. $Z_\odot$ is detected at massive galaxy groups, and we take it as an upper limit \citep{Gastaldello2021}, $0.1Z_\odot$ is reported from the CGM of the Milky-Way \citep{Ponti2023}, and $0.3Z_\odot$ is a commonly used value for the hot CGM. We consider the uncertainties from temperature and metallicity abundance as systematic. The uncertainty of each assumption of temperature and metallicity abundance is estimated by propagating the 1$\sigma$ uncertainty of $\beta$ model considering the covariance of parameters therein. 
 
We calculate the hot CGM mass ($M_{\rm CGM, <R_{\rm vir}}$) by integrating the density profile of the hot CGM within $R_{\rm vir}$. 
For the CEN sample, we sum the hot CGM and stellar mass\footnote{We do not include the satellite stellar mass when calculating the baryon fraction.} $M_{\rm b}=M_{\rm CGM, <R_{\rm vir}}+M_*$, and calculate the baryon fraction by $f_{\rm b, hot\, CGM+star, <R_{\rm vir}}=M_{\rm b}/M_{\rm vir}$. The result of central galaxies with $\log(M_*)=10.5-11.5$ is presented in the left panel of Fig.~\ref{Fig_fbaryon}. For the MW-mass, M31-mass and 2M31-mass galaxies, we obtain $f_{\rm b, hot\, CGM+star, <R_{\rm vir}} = 0.12\pm0.06, 0.07\pm0.02, 0.06\pm0.01$, respectively. We compare our measurements to the $f_{\rm b}$ measured based on the X-ray and tSZ \citep{Li2018,Das2020,Bregman2022,DasChiangMathur_2023ApJ...951..125D,Nicastro2023}. We find consistent results of $f_{\rm b, hot\,CGM+star, <R_{\rm vir}}$ with literature within $1\sigma$. 

For the CEN$_{\rm halo}$ sample, we do not include the stellar component, and the baryon fraction within $R_{\rm vir}$ is defined by $f_{\rm b, hot\,CGM,<R_{vir}}= M_{\rm CGM, <R_{\rm vir}}/M_{\rm vir}$. For the galaxies with $M_{\rm 200m}=12.5-13.0$, $13.0-13.5$ and $13.5-14.0$, we obtain $f_{\rm b, hot\,CGM,<R_{vir}}=0.03\pm0.02$, $0.04\pm0.01$ and $0.05\pm0.005$ for the three halo mass bins respectively. 
We compare our measurements to the $f_{\rm b, hot\,CGM}$ measured within $R_{\rm 500c}$ based on the X-ray emission and tSZ \citep{Sun2009,Sanderson2013,Lovisari2015,Eckert2016,Nugent2020,Akino2022}.
To keep consistency, we also calculate the hot CGM fraction within $R_{\rm 500c}$ ($f_{\rm b, hot\,CGM,<R_{500c}}=M_{\rm CGM, <R_{\rm 500c}}/M_{\rm 500c}$) by integrating the density profile of the hot CGM within $R_{\rm 500c}$\footnote{We convert the $M_{\rm 200m}$ to $M_{\rm 500c}$ with the mass-concentration relation model from \citet{Ishiyama2021} and calculate the $R_{\rm 500c}$.}. The hot CGM fraction within $R_{\rm 500c}$ is about $50\%$ of the hot CGM fraction within $R_{\rm vir}$. The result is presented in the right panel of Fig.~\ref{Fig_fbaryon}. We find our $f_{\rm b, hot\,CGM,<R_{500c}}$ is lower than the literature value, with the difference being approximately $2\sigma$.

The $\Lambda CDM$ cosmology theory predicts $f_{\rm b}\approx16\%$ \citep{Planck2020}, as the dash line in Fig.~\ref{Fig_fbaryon}. Our measured $f_{\rm b, hot\,CGM+star}$ within the virial radius of galaxies are generally below the $\Lambda CDM$ cosmology theory predicted. 
We obtain a hot CGM density profile with a slope $\approx r^{-1.2}$ within $R_{\rm vir}$, which is flatter than the dark matter profile \citep[$\approx r^{-3.0}$, NFW profile,][]{NFW1997}. 
Consequently, as the integrated radius increases, the baryon fraction within that radius also increases. The relatively flat density profile of the hot CGM near the virial radius implies that hot baryons may extend beyond. If the hot CGM density profile can be extrapolated beyond the virial radius without a cutoff, the baryon fraction could reach the cosmological value within a few times $R_{\rm vir}$.

Therefore, We extrapolate the baryon density profile measured within $R_{\rm vir}$ and the NFW profile to larger radii. We calculate the radius within which the baryon fraction exceeds $16\%$ (namely the closure radius $R_{\rm c}$ defined in \citet{Ayromlou2023}\footnote{Notice the $R_{\rm c}$ defined in \citet{Ayromlou2023} does not restrict to the central halo but considers all the dark matter and baryon within $R_{\rm c}$. Differently, we restrict to the central halo by simple extrapolation of its baryon and dark matter profiles, and do not consider the baryon and dark matter contribution from other halos.}), the result is shown in Fig.~\ref{Fig_fbaryon_hist}. We take the baryon fraction derived from the assumption of $T_{\rm vir}$ and $0.3Z_\odot$ to estimate $R_{\rm c}$. We obtain closure radius $R_{\rm c}\approx 3 R_{\rm vir}$ for MW-mass, M31-mass and 2M31-mass galaxies. 

Future, more extensive galaxy surveys, such as DESI and 4MOST, will allow us to expand the galaxy sample and enhance the statistical power necessary to characterize the hot CGM profile beyond the virial radius accurately, therefore allowing us to either probe the presence of the baryons beyond the virial radius or determine the presence of a cut off in the hot CGM density distribution. 

The low baryon fraction within $R_{\rm vir}$ might be due to the ejective feedback processes; for example, AGN outflows might push the baryons out of $R_{\rm vir}$ \citep{Sorini2022, Ayromlou2023}. The SIMBA simulation found $R_{\rm c}\approx 15\,R_{\rm vir}$ for M31-mass galaxies and $R_{\rm c}\approx 10\,R_{\rm vir}$ for 2M31-mass galaxies \citep{Sorini2022}. The IllustrisTNG simulation found $R_{\rm c}\approx 5\,R_{\rm vir}$ and $R_{\rm c}\approx 3\,R_{\rm vir}$, respectively \citep{Ayromlou2023}. The closure radii predicted in the simulations are generally larger than the $3\,R_{\rm vir}$ estimated from the observed and extrapolated hot CGM profile. The X-ray surface brightness profiles measured in this analysis provide a new benchmark for the simulations to implement the feedback processes therein. 

Measuring the baryon mass through the X-ray emission method relies on the accurate assumption of the properties of the hot CGM. Our assumed temperatures and metallicity abundances may not be sufficient to cover the possible physical properties of the hot gas and introduce systematic errors. For example, the temperature of the hot CGM might decrease with radius. In such a case, our assumed isothermal profile overestimates the emissivity of the hot CGM. Besides, we do not consider the other phases of the CGM, namely, the less hot components that are not measured by the X-ray emission. Especially for the lower mass galaxies, the cool phase may occupy a higher share of the gas mass. For example, the SIMBA simulation predicts the gas with $T<0.5T_{\rm vir}$ contributes to about 2\% baryon fraction for MW-mass galaxies \citep{Sorini2022}.

Our result only constitutes a first approximation and does not reflect the wealth of physical processes at stakes in the CGM \citep{Faucher-GiguereOh_2023}. We leave a detailed inference of the different baryonic components and their properties for future studies. With the progress of the metallicity abundance and temperature measurement, point source models, and better statistics from larger galaxy samples in the future, we can set better constraints on the CGM density profile of MW-mass and M31-mass galaxies. In particular, the X-ray microcalorimeter observations, with a high energy resolution, are expected to contribute significantly to this endeavor \citep{athena2013,xrism2020,Cui2020}.

\section{Summary and conclusions}

In this work, we provide a new detection of the X-ray emission from the hot circumgalactic medium (hot CGM) around large statistical samples of (massive) galaxies. 
We use the eRASS:4 X-ray data over an extra-galactic sky of about 11,000 $\rm deg^2$.
We achieved a significant step forward in understanding the influence and possible bias from the satellite galaxies (Figs.~\ref{Fig_sketch},~\ref{Fig_profile_full_nomask} and~\ref{Fig_profile_censat}). 
With the SDSS (spectroscopic) galaxy catalog and the group finder algorithm applied, we built two galaxy samples containing about 85,222 central galaxies split into stellar mass bins and 125,512 central galaxies split in halo mass bins (Table~\ref{Tab:gal:samples}). 
We carefully model the components that can contaminate the CGM emission: the unresolved AGN and XRB in the stacked galaxies, the XRB in the satellite galaxies, and the satellite boost bias due to misclassified centrals (Sect.~\ref{Sec_model}). 

We stacked the galaxies, and the statistics we gained support the following main findings:

   \begin{itemize}
      \item We measured the mean X-ray surface brightness profiles for central galaxies with $\log(M_*)>10.0$ or $\log(M_{\rm 200m})>11.5$ to unprecedentedly high S/N values (Fig.~\ref{Fig_profile_ms}).
      We detected extended X-ray emission around $\log(M_*)>10.5$ and $\log(M_{\rm 200m})>12.5$ galaxy populations out to $R_{\rm vir}$. The S/Ns of the extended emission within $R_{\rm vir}$ are $>3.1\sigma$ for $\log(M_*)>10.5$ central galaxies and $>2.8\sigma$ for central galaxies with $\log(M_{\rm 200m})>12.5$.
      \item For the first time, we have measured the mean X-ray surface brightness profile of hot CGM for approximately volume-limited central $\log(M_*)=10.5-11.5$ or $\log(M_{\rm 200m})=12.5-14.0$ galaxy populations out to $R_{\rm vir}$ (Fig.~\ref{Fig_profile_cgm} and Fig.~\ref{Fig_profile_halo_cgm}). We fit a $\beta$ model to the profiles and obtained $\beta\approx0.4$ (Table~\ref{tab:betamodel}). The measured X-ray surface brightness profiles are essential to constraining the galaxy evolution models in semi-analytic or hydrodynamic simulations.
      \item We conservatively assumed a range of temperatures and metallicity abundances of the hot CGM to infer its mass. 
      The baryon fraction within $R_{\rm vir}$ is lower than the $\Lambda CDM$ cosmology theory predicted and is consistent with literature measurements (Fig.~\ref{Fig_fbaryon}). We extrapolate the measured hot CGM profile beyond $R_{\rm vir}$ and find within $\approx 3 R_{\rm vir}$ the baryon fraction is close to the $\Lambda CDM$ cosmology theory (Fig.~\ref{Fig_fbaryon_hist}). 
      
   \end{itemize}

The eROSITA sky survey yields a vast repository of X-ray data across the sky, with abundant information on the hot CGM that calls for further investigation and analysis. 
The measured X-ray surface brightness profiles of the FULL$_{\rm phot}$ have high S/N values, which allows for a complete model to be built, including the galaxy distribution and X-ray properties of the galaxy (Shreeram et al. in prep).
With deeper galaxy surveys (e.g., 4MOST, DESI) we can enlarge the galaxy sample, increase the statistics, and obtain a better view of the hot CGM around dwarf galaxies ($\log(M_*)<10$). With the continued progress of X-ray microcalorimeter observations, we will obtain a better knowledge of the metallicity and temperature of the CGM to derive its density profile.

\begin{acknowledgements}
We thank the anonymous referee for thoughtful comments that improved the manuscript.
This project acknowledges financial support from the European Research Council (ERC) under the European Union's Horizon 2020 research and innovation program HotMilk (grant agreement No. 865637). GP acknowledges support from Bando per il Finanziamento della Ricerca Fondamentale 2022 dell'Istituto Nazionale di Astrofisica (INAF): GO Large program and from the Framework per l'Attrazione e il Rafforzamento delle Eccellenze (FARE) per la ricerca in Italia (R20L5S39T9). NT acknowledges support from NASA under award number 80GSFC21M0002. PP has received funding from the European Research Council (ERC) under the European Union's Horizon Europe research and innovation program ERC CoG (Grant agreement No. 101045437).\\
This work is based on data from eROSITA, the soft X-ray instrument aboard SRG, a joint Russian-German science mission supported by the Russian Space Agency (Roskosmos), in the interests of the Russian Academy of Sciences represented by its Space Research Institute (IKI), and the Deutsches Zentrum für Luft- und Raumfahrt (DLR). The SRG spacecraft was built by Lavochkin Association (NPOL) and its subcontractors, and is operated by NPOL with support from the Max Planck Institute for Extraterrestrial Physics (MPE).\\

The development and construction of the eROSITA X-ray instrument was led by MPE, with contributions from the Dr. Karl Remeis Observatory Bamberg \& ECAP (FAU Erlangen-Nuernberg), the University of Hamburg Observatory, the Leibniz Institute for Astrophysics Potsdam (AIP), and the Institute for Astronomy and Astrophysics of the University of Tübingen, with the support of DLR and the Max Planck Society. The Argelander Institute for Astronomy of the University of Bonn and the Ludwig Maximilians Universität Munich also participated in the science preparation for eROSITA.\\
The eROSITA data shown here were processed using the eSASS/NRTA software system developed by the German eROSITA consortium. \\
Funding for the SDSS and SDSS-II has been provided by the Alfred P. Sloan Foundation, the Participating Institutions, the National Science Foundation, the U.S. Department of Energy, the National Aeronautics and Space Administration, the Japanese Monbukagakusho, the Max Planck Society, and the Higher Education Funding Council for England. The SDSS Web Site is http://www.sdss.org/.

The SDSS is managed by the Astrophysical Research Consortium for the Participating Institutions. The Participating Institutions are the American Museum of Natural History, Astrophysical Institute Potsdam, University of Basel, University of Cambridge, Case Western Reserve University, University of Chicago, Drexel University, Fermilab, the Institute for Advanced Study, the Japan Participation Group, Johns Hopkins University, the Joint Institute for Nuclear Astrophysics, the Kavli Institute for Particle Astrophysics and Cosmology, the Korean Scientist Group, the Chinese Academy of Sciences (LAMOST), Los Alamos National Laboratory, the Max-Planck-Institute for Astronomy (MPIA), the Max-Planck-Institute for Astrophysics (MPA), New Mexico State University, Ohio State University, University of Pittsburgh, University of Portsmouth, Princeton University, the United States Naval Observatory, and the University of Washington.
\\
The DESI Legacy Imaging Surveys consist of three individual and complementary projects: the Dark Energy Camera Legacy Survey (DECaLS), the Beijing-Arizona Sky Survey (BASS), and the Mayall z-band Legacy Survey (MzLS). DECaLS, BASS and MzLS together include data obtained, respectively, at the Blanco telescope, Cerro Tololo Inter-American Observatory, NSF’s NOIRLab; the Bok telescope, Steward Observatory, University of Arizona; and the Mayall telescope, Kitt Peak National Observatory, NOIRLab. NOIRLab is operated by the Association of Universities for Research in Astronomy (AURA) under a cooperative agreement with the National Science Foundation. Pipeline processing and analyses of the data were supported by NOIRLab and the Lawrence Berkeley National Laboratory (LBNL). Legacy Surveys also uses data products from the Near-Earth Object Wide-field Infrared Survey Explorer (NEOWISE), a project of the Jet Propulsion Laboratory/California Institute of Technology, funded by the National Aeronautics and Space Administration. Legacy Surveys was supported by: the Director, Office of Science, Office of High Energy Physics of the U.S. Department of Energy; the National Energy Research Scientific Computing Center, a DOE Office of Science User Facility; the U.S. National Science Foundation, Division of Astronomical Sciences; the National Astronomical Observatories of China, the Chinese Academy of Sciences and the Chinese National Natural Science Foundation. LBNL is managed by the Regents of the University of California under contract to the U.S. Department of Energy. The complete acknowledgments can be found at https://www.legacysurvey.org/acknowledgment/.\\
The Siena Galaxy Atlas was made possible by funding support from the U.S. Department of Energy, Office of Science, Office of High Energy Physics under Award Number DE-SC0020086 and from the National Science Foundation under grant AST-1616414.\\

\end{acknowledgements}

\bibliographystyle{aa}
\bibliography{ref.bib}

\begin{appendix} 

\section{Validation of the background calculation}\label{Apd_bg}

\begin{figure*}
\centering
\includegraphics[width=0.32\linewidth]{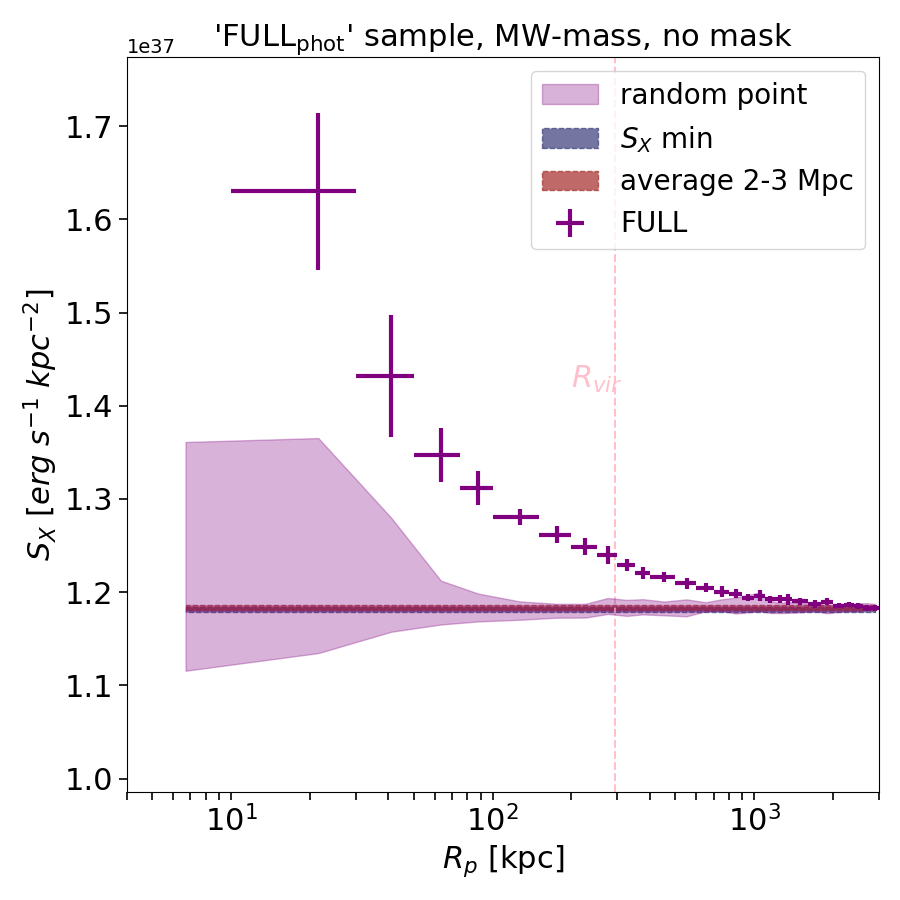}
\includegraphics[width=0.32\linewidth]{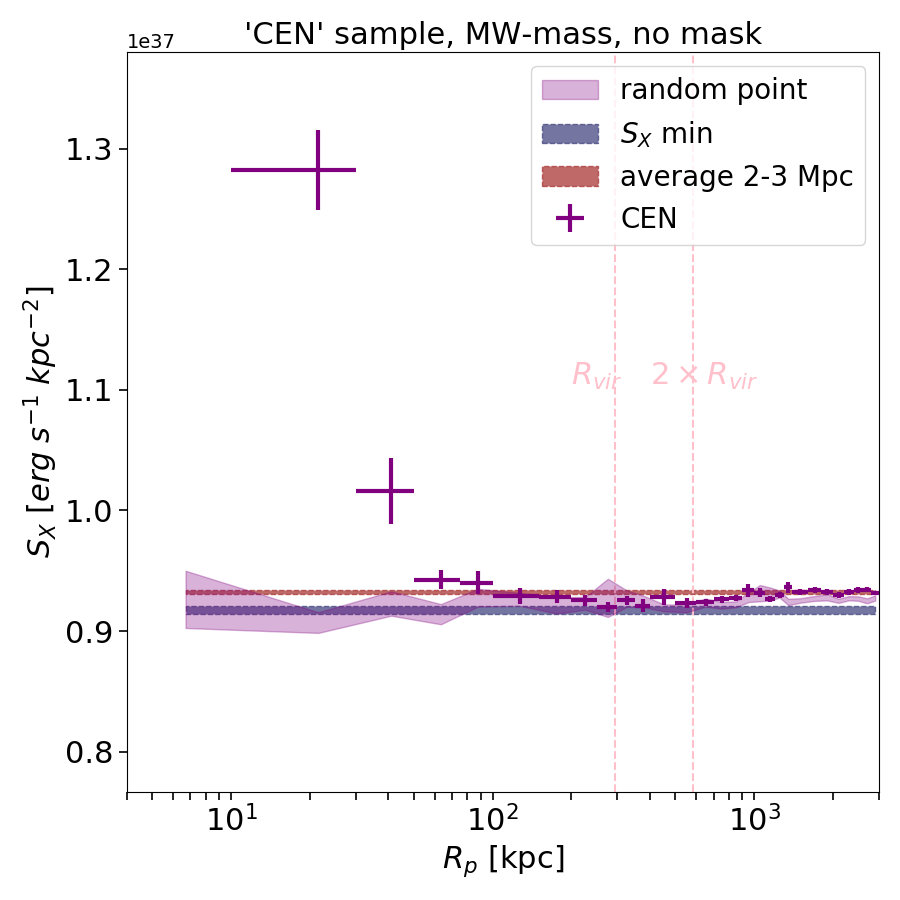}
\includegraphics[width=0.32\linewidth]{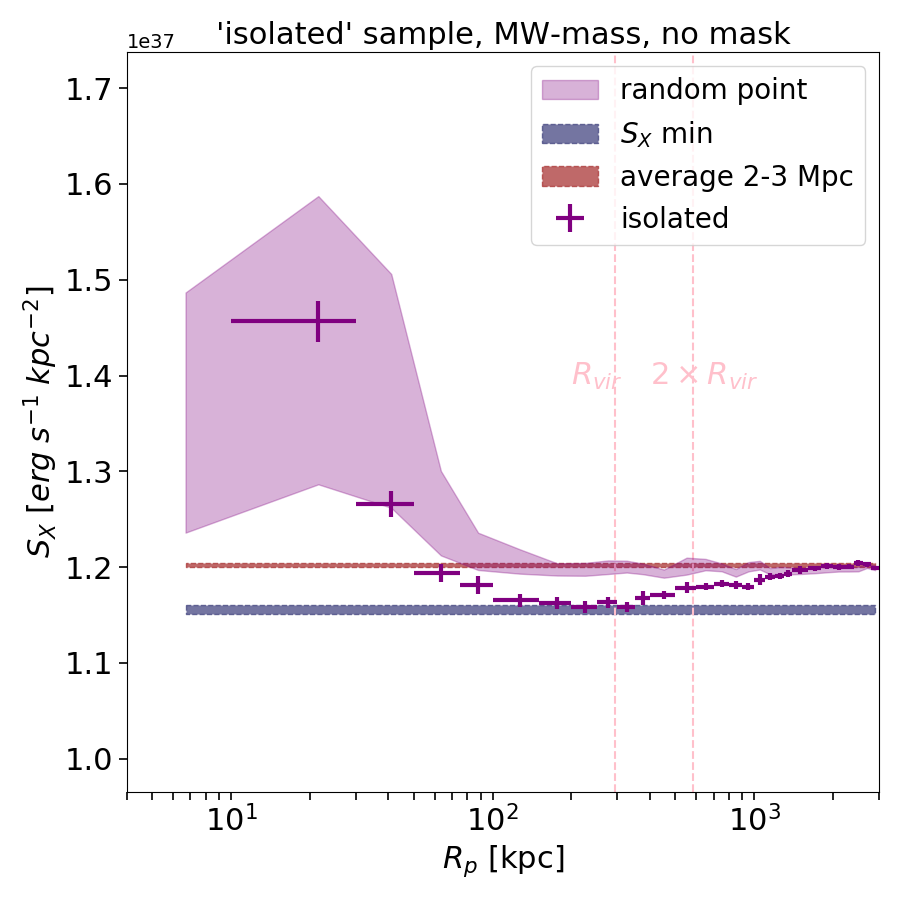}
\caption{Observed X-ray surface brightness profiles of FULL$_{\rm phot}$ sample (left), CEN sample (middle) and isolated sample (right), without masking X-ray sources. The data points are the observed X-ray surface brightness profile, the colored bands are the estimated background by taking the random points (purple), the minimum X-ray surface brightness (blue) and the mean X-ray surface brightness within 2-3 Mpc (red).} 
\label{Fig_bg}
\end{figure*}

We test different methods to estimate the average background for the stacked galaxies. 

Method a) We retrieve events within 3 Mpc of a random point which is located within the $1-4\times R_{\rm vir}$ of the galaxy ($\rm CUBE_{rd}$). We merge $\rm CUBE_{rd}$ of each galaxy sample and calculate the mean $S_{\rm X,rd}$ within 3 Mpc. 
As $\rm CUBE_{rd}$ is randomly distributed over the sky area of stacked galaxies, $S_{\rm X,rd}$ is the average X-ray background of the stacked area. 

Method b) We estimate the background surface brightness ($S_{\rm X,Mpc}$) by taking the mean $S_{\rm X,gal}$ within the annulus at 2-3 Mpc distance from the center. The annulus choice of the $S_{\rm X,Mpc}$ guarantees that it is well beyond the virial radius of the largest haloes, namely the galaxy clusters.
Method c) We define $S_{\rm X,min}$ as the minimum value of $S_{\rm X,gal}$ between $R_{\rm 500c}$ and $2\times R_{\rm vir}$. 

The X-ray surface brightness profiles of MW-mass galaxies in the FULL$_{\rm phot}$, isolated and CEN samples compared to the different background estimates are presented in Fig.~\ref{Fig_bg}. Beyond about 2 Mpc, the X-ray surface brightness profiles converge to $S_{\rm X,rd}$ (method a), meaning the average background is properly recovered by the methods.
For the FULL$_{\rm phot}$ sample, the difference between the three background estimates is less than 0.3\% (at all radii), which is less than the uncertainty of each background. Therefore, the method used to estimate the background does not influence the result, and we take $S_{\rm X, min}$ as the background estimation for FULL$_{\rm phot}$ (or FULL$_{\rm spec}$). 

The environment in which galaxies live may cause the local X-ray background to differ from the average background. 
In particular, the X-ray background towards isolated galaxies is expected (and observed) to be lower than the average, as they are selected in projection to live in sparse environments and have fewer foreground and background galaxies contributing to the X-ray background along that line of sight. 
Central galaxies, defined as the most massive galaxies in their local environment (virial halo), are expected to have only lower-mass foreground and background galaxies within their virial halo projection. Therefore, along the line of sight towards central galaxies, the local X-ray background is also lower than average but this effect is less severe than for isolated galaxies.

For the isolated sample, the observed X-ray surface brightness decreases up to $R_{\rm vir}$ and then increases until reaching the $S_{\rm X,rd}$ at 2Mpc, as we discussed above (also see isolated galaxy definition in Appendix~\ref{Sec_iso}). The difference between $S_{\rm X,min}$ and $S_{\rm X,rd}$ is about $4.1\% S_{\rm X,rd}$.
We take $S_{\rm X,min}$ as the proper local background estimate for the isolated galaxy sample. 
The X-ray surface brightness of the CEN sample decreases until $1-2\ R_{\rm vir}$, then increases to the $S_{\rm X,rd}$ (agrees within 1$\sigma$) but to a lesser extent than the isolated sample. 
The difference between the three background estimates is about $1.5\% S_{\rm X,rd}$. 
We also take $S_{\rm X,min}$ as the proper local background of the central galaxy samples (CEN and CEN$_{\rm halo}$). We use the satellite galaxy samples (SAT and SAT$_{\rm halo}$) to model the contamination from the misclassified centrals in the CEN and CEN$_{\rm halo}$ samples. To keep consistency, we also take $S_{\rm X, min}$ as the background for SAT and SAT$_{\rm halo}$.
In summary, we take the $S_{\rm X,min}$ as the proper background in all of our samples.

The background surface brightness is redshift-dependent in units of $\rm erg\,s^{-1}\,kpc^{-2}$. The $S_{\rm X,rd}$ background levels of the isolated and CEN samples are therefore different by a factor of about $(\frac{1+z_{\rm iso}}{1+z_{\rm cen}})^5$, where $z_{\rm iso}$ and $z_{\rm cen}$ are the median redshifts of the isolated and `CEN' sample. 

\section{Merge SGA catalog to LS DR9 catalog}\label{Apd_catalog}
The SGA is a multi-wavelength atlas of nearby galaxies selected from DESI Legacy Imaging Surveys DR9. The selection is based on the apparent angular diameter of galaxies that are large enough to be spatially resolved. In the western Galactic hemisphere, there are 194527 galaxies in the SGA catalog. We match SGA to the LS DR9 catalog and find about 15\% galaxies in SGA have no match in LS DR9. We add the missing SGA galaxies to the LS DR9 catalog to avoid losing the nearby massive galaxies.

No $M_*$ information is provided in SGA, and about 34\% of SGA galaxies have no redshift estimation. For SGA galaxies with a redshift estimate, we estimate the stellar mass by the SED fitting using the three optical bands (g,r,z) and two infrared bands (W1, W2), following the method in \citet{Zou2019}. The galaxies without redshift estimation are thrown away due to the difficulty of estimating stellar mass. There are about 5\% galaxies that fall in the stacking area with no photometry information and low-quality redshift and $M_*$, which we throw away. 
We match the remaining SGA galaxies to GAMA and find that our derived redshift and $M_*$ are good quality. 
We add the 15,735 missing SGA galaxies with $\log(M_*)>9$ to the LS DR9 catalog. 90\% of these galaxies have $z<0.1$. 

Besides, galaxies in the LS DR9 catalog could be too close to the bright nearby SGA galaxy. This may affect the accuracy of $z_{\rm phot}$ or $M_*$ in the LS DR9 catalog. We mask the LS DR9 galaxies that are within the $R_{25}$ of SGA galaxies, where $R_{25}$ is the approximate major-axis radius at the $25\,\rm mag\ arcsec^{-2}$, which can be taken as the radius of galaxy disk. Notice that we did not consider the inclination and position angle of the SGA galaxy. The procedure above removes $0.01\%$ of LS DR9 galaxies.

\section{AGN model for CEN sample}\label{Apd_agn}
\begin{table}[h]
\begin{center}
\caption{Percentage and number of galaxy hosting AGN in CEN and CEN$_{\rm halo}$ sample, identified using the BPT diagram \citep{BPT1981,Brichmann2004}. }
\label{Tab_cenagn}
\begin{tabular}{ccccccccc} 
\hline
\hline
$\log(M_*)$ &$f_{\rm AGN}$ &$N_{\rm CENAGN}$\\
\hline
10.0-10.5&16\%&1261\\
10.5-11.0&18\%&5421\\
11.0-11.25&14\%&3447\\
11.25-11.5&8\%&1505\\
\hline
\end{tabular}

\begin{tabular}{ccccccccc} 
$\log(M_{\rm 200m})$ &$f_{\rm AGN}$ &$N_{\rm CENAGN}$\\
\hline
11.5-12.0&9\%&2396\\
12.0-12.5&17\%&7113\\
12.5-13.0&13\%&3965\\
13.0-13.5&4\%&849\\
13.5-14.0&2\%&170\\
\hline
\end{tabular}
\end{center}
\end{table}

\begin{figure*}
\centering
\includegraphics[width=0.48\linewidth]{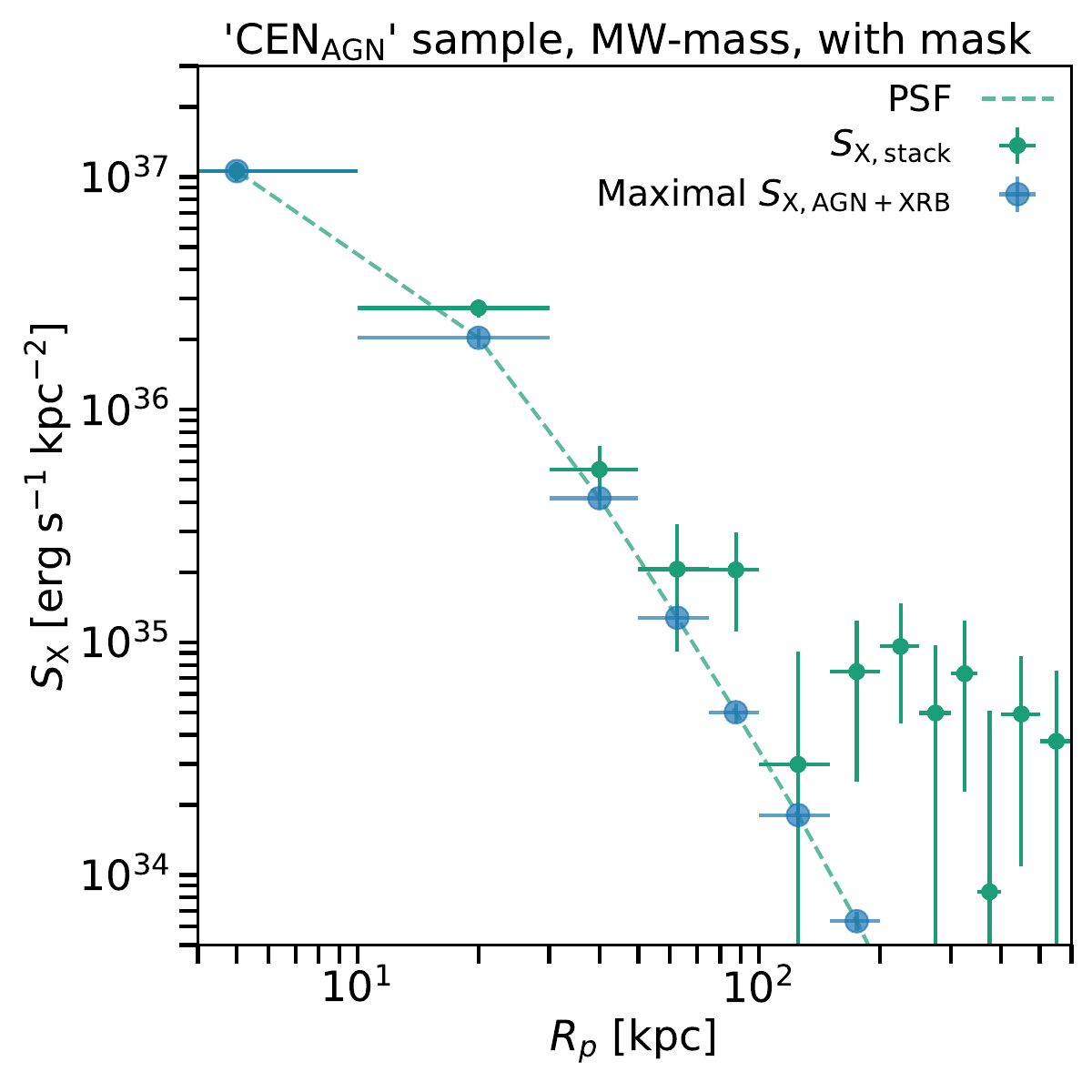}
\includegraphics[width=0.48\linewidth]{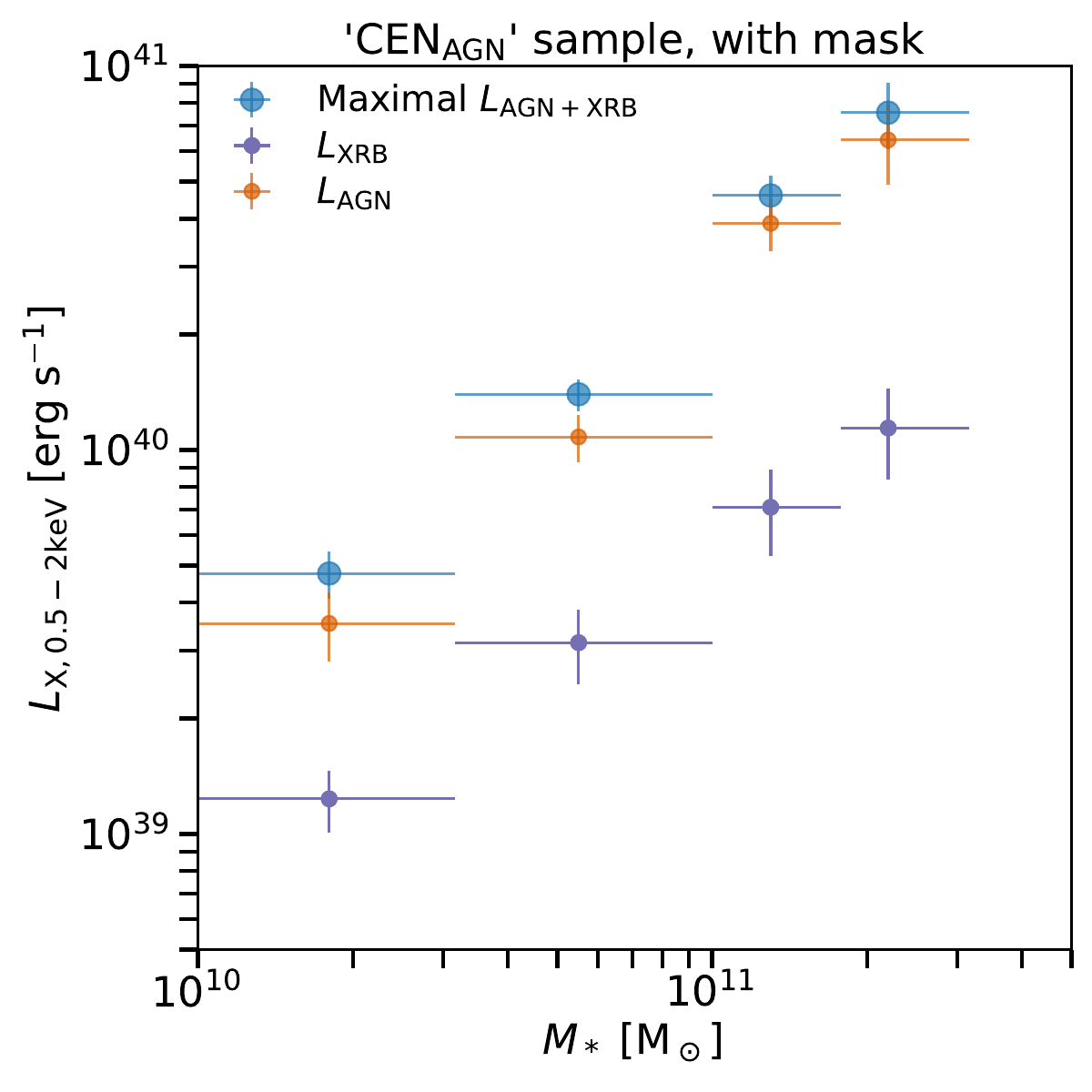}

\caption{X-ray surface brightness profile of MW-mass galaxies (left) in CEN$_{\rm AGN}$ sample, after subtracting background and masking X-ray sources (green points), the maximal X-ray emission from the unresolved point sources (blue points), in the shape of PSF (green dashed line). Observed maximal X-ray luminosity (right) of unresolved point source (blue), the model predicted XRB luminosity (purple), and the residual X-ray luminosity of unresolved AGN (orange). } 
\label{Fig_agn_cen}
\end{figure*}

We estimate the AGN contamination in the CEN sample based on observations. 
We select the central galaxies hosting AGN, identified as AGN or composite, based on the BPT diagram, and name it as CEN$_{\rm AGN}$ \citep{BPT1981,Brichmann2004}. The percentage of central galaxies identified as hosting AGN ($f_{\rm AGN}$) is listed in Table~\ref{Tab_cenagn}. We stack the CEN$_{\rm AGN}$ sample in the same bins as the CEN sample, and mask detected X-ray sources.
The obtained X-ray surface brightness profile for the MW-mass CEN$_{\rm AGN}$ galaxies is plotted in the left panel of Fig.~\ref{Fig_agn_cen}.
The maximal X-ray emission originating from the unresolved point sources (AGN and XRB) in the stacked galaxies should be in the shape of PSF, with the first bin as bright as detected (green dots).
The maximal X-ray luminosity of the unresolved point sources is calculated by integrating the X-ray emission in the shape of PSF out to $R_{\rm 500c}$.
We subtract the XRB luminosity predicted by the empirical model ($L_{\rm XRB}$) \citep{Aird2017}, and obtain the maximal unresolved AGN emission in the stacked galaxies ($L_{\rm AGN}$), see right panel of Fig.~\ref{Fig_agn_cen}. 
For the galaxies identified as hosting AGN, the unresolved AGN X-ray emission is about $3-5$ times brighter than the XRB. 
We notice that only $<20\%$ galaxies in the CEN or CEN$_{\rm halo}$ samples are identified as hosting AGNs; we summarize that after averaging, the unresolved AGN emission is comparable or subdominant compared to the XRBs for the CEN or CEN$_{\rm halo}$ samples.

\section{Isolated galaxy sample and stacking}\label{Sec_iso}
\subsection{Build isolated galaxy sample from LS DR9}
\begin{figure}
        \centering
        \includegraphics[width=0.97\columnwidth]{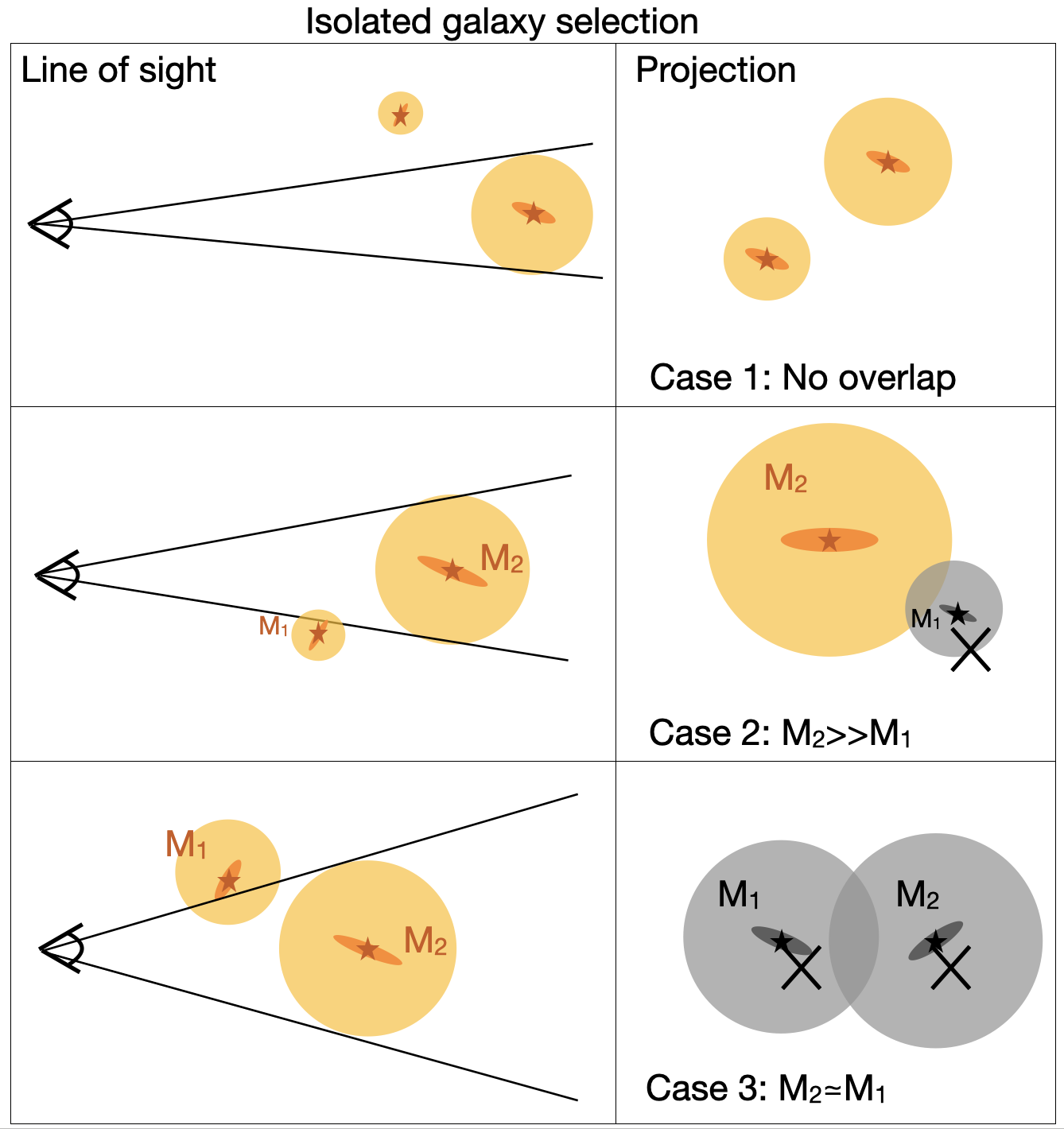}
        \caption{Illustration of how the isolated selection works for photometric samples. If after projection, the virial halo of a galaxy has no overlap or is much more massive than overlapped galaxies, they are kept in the isolated galaxy sample (in orange color). Otherwise, they are removed (in color gray). The radius of the largest circle denotes the virial radius of the galaxy (not to scale).
        }
        \label{Fig_sketch_iso}
\end{figure}

Unlike spectroscopic surveys such as SDSS, the central galaxy selection algorithm by group finder is not readily applicable to LS DR9 due to its photometric redshift uncertainties. 
Instead, we construct an `isolated' galaxy sample from LS DR9 that is, as much as possible, free from satellite contamination. 

We work with projected haloes, as we illustrate schematically in Fig.~\ref{Fig_sketch_iso}.
For each galaxy, based on its stellar mass, we compute its mean virial radius ($R_{\rm vir}$) and the angle it subtends on the sky ($\theta_{\rm vir}$) using the stellar to halo mass relation from \citet{Moster2010}. The virial radius definition is that of \citet{Bryan1998}.
In the simplest case (case 1), if the virial halo of a galaxy has no overlap with other galaxy halos along the line of sight (i.e., the angular separation of a galaxy pair is larger than the sum of their virial radii), then we assume it is naturally central and does not suffer from satellite contamination.
In the case where the virial halo of a galaxy overlaps with other galaxy halos along the line of the sight (case 2, case 3), we select the galaxy that is dominantly more massive (by a factor of two) and remove the smaller galaxy from the sample (case 2). If the two galaxies have comparable stellar masses (the stellar mass difference is smaller than two), we reject both galaxies. In this way, we get the central galaxy at least twice as massive as the most massive satellite galaxy in its halo. It means that the satellite boost and the contamination from the sources in satellite galaxies are reduced. The sample is a (incomplete) sub-sample of the central galaxy population living in under-dense environments. The X-ray background for the isolated galaxies is lower than the average X-ray background over the sky (due to the selection method of isolated galaxies), and this necessitates taking $S_{\rm X,min}$ as the background (see the discussion in Appendix~\ref{Apd_bg}). The selection of isolated galaxies has a low dependence on the photometric redshift accuracy of the galaxies. 

The selection method used to construct the isolated galaxy sample is validated through the mock galaxy catalog. 
We build a mock LS DR9 catalog from the light cone we build in Sect.~\ref{Sec_mockcat} up to z=0.45. We apply the magnitude cut $r_{AB}<21$ to the light cone. We add a Gaussian distributed $z_{\rm error}$ and $M_{\rm error}$ with the same median and dispersion of the LS DR9 catalog to the corresponding accurate values provided by the light cone. 
Then we select the galaxies within the same $M_*$ and $z_{\rm phot}$ bins as FULL$_{\rm phot}$ to build the mock FULL$_{\rm phot}$ sample. We apply the isolated galaxy selection scheme to the mock FULL$_{\rm phot}$ sample and obtain a mock isolated sample.
We use the mock samples to assess the isolated selection effect of our galaxy samples.
In the mock FULL$_{\rm phot}$ sample, we find that the satellite fraction amounts to 25-35\%. 
In the mock isolated sample, we get much lower satellite fractions, only 5-10\%. In most cases, the satellite galaxies are kept since they are far away from other massive galaxies. A small portion of the satellite galaxies are misidentified due to the dispersion of $Z_{\rm phot}$ and $M_*$ of LS DR9. 
We also use the mock LS DR9 catalog to derive the stacked galaxies' average underlying $M_{\rm halo}$ and $R_{\rm vir}$. 

\subsubsection*{Isolated galaxy sample (isolated)}

Following the scheme defined above, we select the isolated galaxies from the FULL$_{\rm phot}$ sample to minimize the projected X-ray emission from nearby galaxies. 
Considering the dispersion of the $z_{\rm phot}$ and $M_*$ of LS DR9, we increase the `FULL$_{\rm phot}$' galaxy sample by retrieving galaxies with $z_{\rm phot}<z_{\rm max}+0.05$ and $M_*>1/2M_{\rm min}$ to avoid extra selection of isolated galaxy. 

The fraction of isolated galaxies selected from the FULL$_{\rm phot}$ sample is 25\% at the lower mass end, 17\% for MW-mass bin, 8\% for M31-mass bin, and 5\% for 2M31-mass bin.
Table~\ref{Tab:gal:samples} details the total number of isolated galaxies obtained for each sample. We name the isolated galaxy sample as `isolated'. In total, we have 213,514 isolated galaxies, about 2.5 times the CEN sample size, about 13 times larger than \citet{Comparat2022}, 130 times larger than \citet{Chada2022} and close to the sample size of \citet{Anderson2015}.

Two main differences exist between this isolated sample and the CEN sample. First, the isolated galaxies are central galaxies with at most $M_{\rm sat}=1/2 M_{\rm cen}$ satellites in the halo, and the satellite boost is well removed. Therefore, the isolated galaxies are located in a less dense environment than the CEN sample and should be considered as a subsample of the central galaxy population.
Second, the redshift distributions of the isolated and CEN samples are different, with the isolated sample extending to $z=0.4$. 

\subsection{The X-ray surface brightness profiles of isolated sample}
\begin{figure}
    \centering
    \includegraphics[width=1.0\columnwidth]{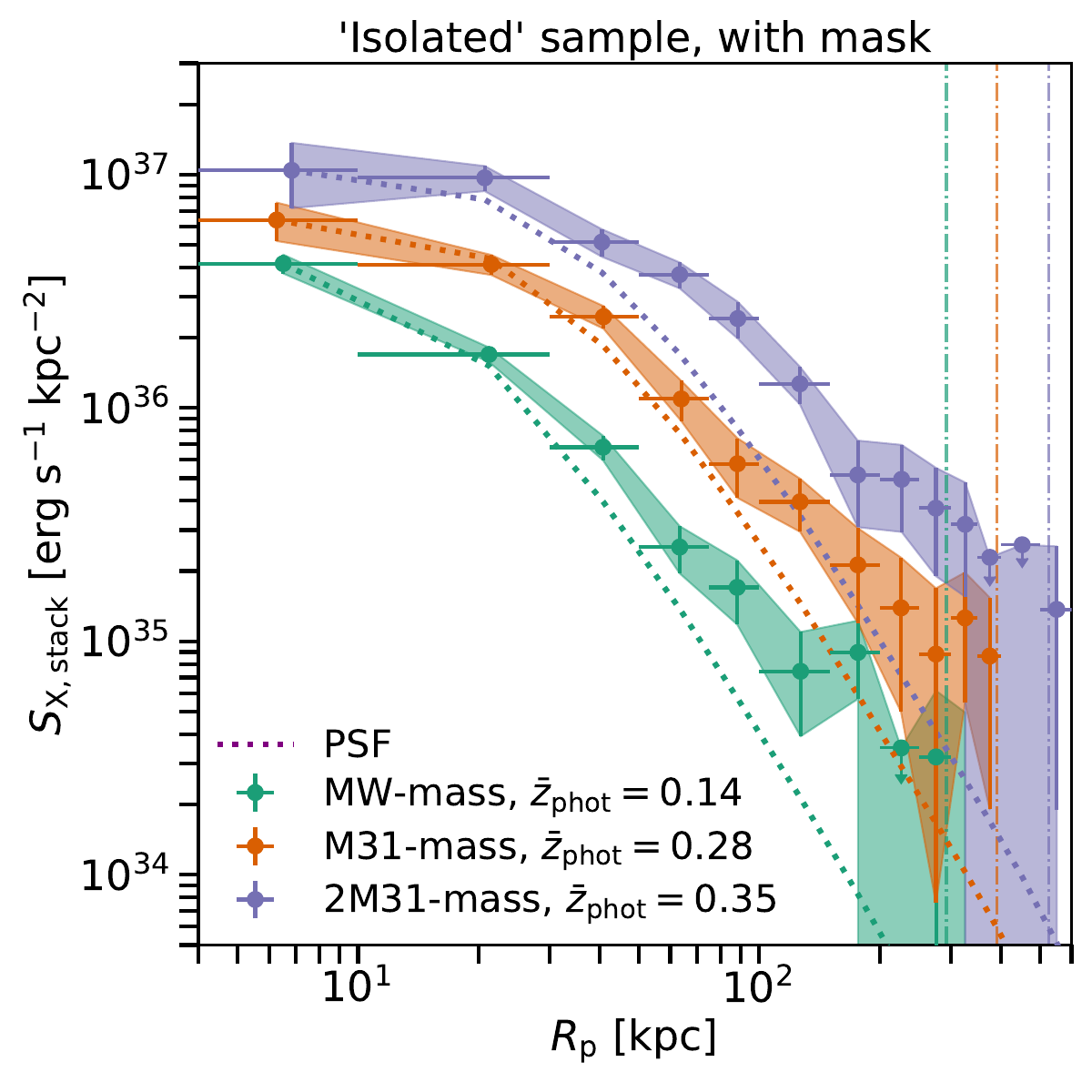}

    \caption{Mean X-ray surface brightness profiles of the isolated sample (with a magnitude limit of $r_{\rm AB}<20$) after subtracting background and masking X-ray sources, for galaxies with $\log(M_*)=10.5-11.0$ (MW-mass), 11.0-11.25 (M31-mass) and 11.25-11.5 (2M31-mass). The X-ray emission comprises unresolved point sources and hot gas. $\Bar{z}$ is the median redshift of galaxies in each mass bin. The dotted lines are the PSF of eROSITA. The vertical dash-dotted lines are the virial radius at different stellar mass bins correspondingly.
    }
        \label{Fig_profile_ms_iso}
\end{figure}

The X-ray surface brightness profiles up to $R_{\rm vir}$
of the isolated sample in MW-mass, M31-mass and 2M31-mass are plotted in Fig.~\ref{Fig_profile_ms_iso}. 
The X-ray emission detected here is the sum of unresolved point sources and hot gas emission. The X-ray surface brightness increases with stellar mass. Extended X-ray emission is detected beyond about 50 kpc for MW-mass and more massive galaxies. The profile becomes more extended with increasing stellar mass. 
The X-ray emission is detected within the virial radius of galaxies with $S/N \approx 3.0,4.9,3.7$ for the MW-mass, M31-mass, and 2M31-mass galaxies in the isolated sample.

\end{appendix}

\end{document}